\documentclass[aps,twocolumn,prl,amsmath,amssymb,10pt,aps]{revtex4-1} 
\usepackage{graphicx}
\usepackage{latexsym}
\usepackage{amsmath}
\usepackage{amsthm}
\usepackage{amsbsy}
\usepackage{amssymb}
\usepackage{epstopdf} 
\usepackage{enumerate}
\usepackage{setspace} 
\usepackage{dcolumn}
\usepackage{bm}
\usepackage{setspace} 
\usepackage{slashed}
\usepackage{color}
\usepackage{youngtab}
\usepackage[colorlinks=true]{hyperref}
\usepackage{float}
\usepackage{graphicx}
\usepackage{multirow}
\usepackage{array}
\newcolumntype{L}[1]{>{\raggedright\arraybackslash}p{#1}}
\newcolumntype{C}[1]{>{\centering\arraybackslash}p{#1}}
\newcolumntype{R}[1]{>{\raggedleft\arraybackslash}p{#1}}

\usepackage{color}
\usepackage{makecell}

\begin{document}
\title{Instability of $j= 3/2$ Bogoliubov Fermi-surfaces }
 
 \author{ Hanbit Oh }
 \author{ Eun-Gook Moon }
 \thanks{egmoon@kaist.ac.kr}
\affiliation{Department of Physics, Korea Advanced Institute of Science and Technology (KAIST), Daejeon 34141, Korea}

\date{\today}
\begin{abstract}
Exotic quantum phases including  topological states and non-Fermi liquids may be realized by quantum states with total angular momentum $j=3/2$, as manifested in HgTe and pyrochlore iridates. Recently, an exotic superconducting state with non-zero density of states of zero energy Bogoliubov (BG) quasiparticles, Bogoliubov Fermi-surface (BG-FS), was also proposed in a centrosymmetric $j=3/2$ system, protected by a Z$_2$ topological invariant. 
 Here, we consider interaction effects of a centrosymmetric BG-FS and demonstrate its instability by using mean-field and renormalization group analysis. 
 The Bardeen-Cooper-Schrieffer (BCS) type logarithmical enhancement is shown in fluctuation channels associated with inversion symmetry. 
 Thus, we claim that the inversion symmetry instability is an intrinsic characteristic of a BG-FS under generic attractive interactions between BG quasiparticles. 
In drastic contrast to the standard BCS superconductivity, a Fermi-surface may generically survive even with the instability.
 We propose the experimental setup, a second harmonic generation experiment with a strain gradient, to detect the instability. Possible applications to iron based superconductors and heavy fermion systems including FeSe are also discussed. 
\end{abstract}
 
\maketitle 

{\it Introduction} : 
Electrons on a lattice may form quantum states with a total angular momentum $j=3/2$, especially with strong spin-orbit coupling \cite{Witczak2}. Cubic and time reversal symmetries may protect degeneracy of the the four states as in GaAs and HgTe. A minimal model of the  $j=3/2$ band structures were provided by Luttinger, so-called Luttinger Hamiltonian, \cite{Luttinger,Murakami} and its low energy properties have been thoroughly understood, being a backbone of semiconductor physics \cite{Yu}.  

Recent advances in topology and correlation research unveil unconventional phases associated with the Luttinger Hamiltonian. 
Topological insulators may be realized by breaking cubic symmetry, for example applying uniaxial pressure, \cite{TI_Inv, Hasan} and Weyl semi-metals may be formed by breaking time reversal symmetry, for example the onset of all-in-all-out order parameter in pyrochlore iridates \cite{Savrasov, Armitage}. In the presence of the long range Coulomb interaction, either non-Fermi liquid or topological states with broken symmetries may be realized with renormalized physical quantities \cite{ Moon_NFL}, and significant advances in experiments have been reported recently \cite{Pr2Ir2O7,Kondo, Nd2Ir2O7, Tokura}.  Furthermore,  quantum phase transitions between the unconventional phases have been investigated finding new universality classes \cite{Herbut,Roy2,Savary,Han3,Hanbit}.   

Exotic superconducting states with $j=3/2$ states were also proposed  \cite{Agterberg1,Agterberg2,Agterberg3,Venderbos,Savary2,Roy,Nakajima,Kim,Igor,GiBaik,GiBaik3,Witczak4,Timm,Brydon,Ghorashi}. In the presence of inversion symmetry, it was proven that a non-interacting Bougoliubov Hamiltonian may host a Fermi-surface of Bougoliubov quasi-particles in drastic contrast to standard nodeless, point node, and line node gap structures, named Bogoliubov Fermi-surface. It is characterized by a Z$_2$ topological invariant of the  Hamiltonian \cite{Agterberg1,Youichi}, and several heavy fermion systems such as URu$_2$Si$_2$ and UBe$_{13}$ are suggested as candidate material though its presence has not been reported yet \cite{Schemm,Matsuda,Heffner,Zieve}.  
It is highly desired to uncover characteristics of a BG-FS for its discovery. 

In this work, we propose enhanced fluctuations of an inversion symmetry order parameter as a key property of a centrosymmetric BG-FS. 
It is shown that a centrosymmetric BG-FS becomes unstable at zero temperature under infinitesimally weak interaction between BG quasi-particles, and the Fermi-surface manifold may be changed, as illustrated in Fig \ref{F1}.     
Our results indicate that an inversion order parameter must be included in a phenomenological Ginzburg-Landau functional of a centrosymmetric BG-FS, and we also propose  second harmonic generation experiments with strain gradient to identify enhanced fluctuations of an inversion symmetry order parameter.   
\begin{figure}[tb]
\centering
\ \ 
\includegraphics[scale=0.26]{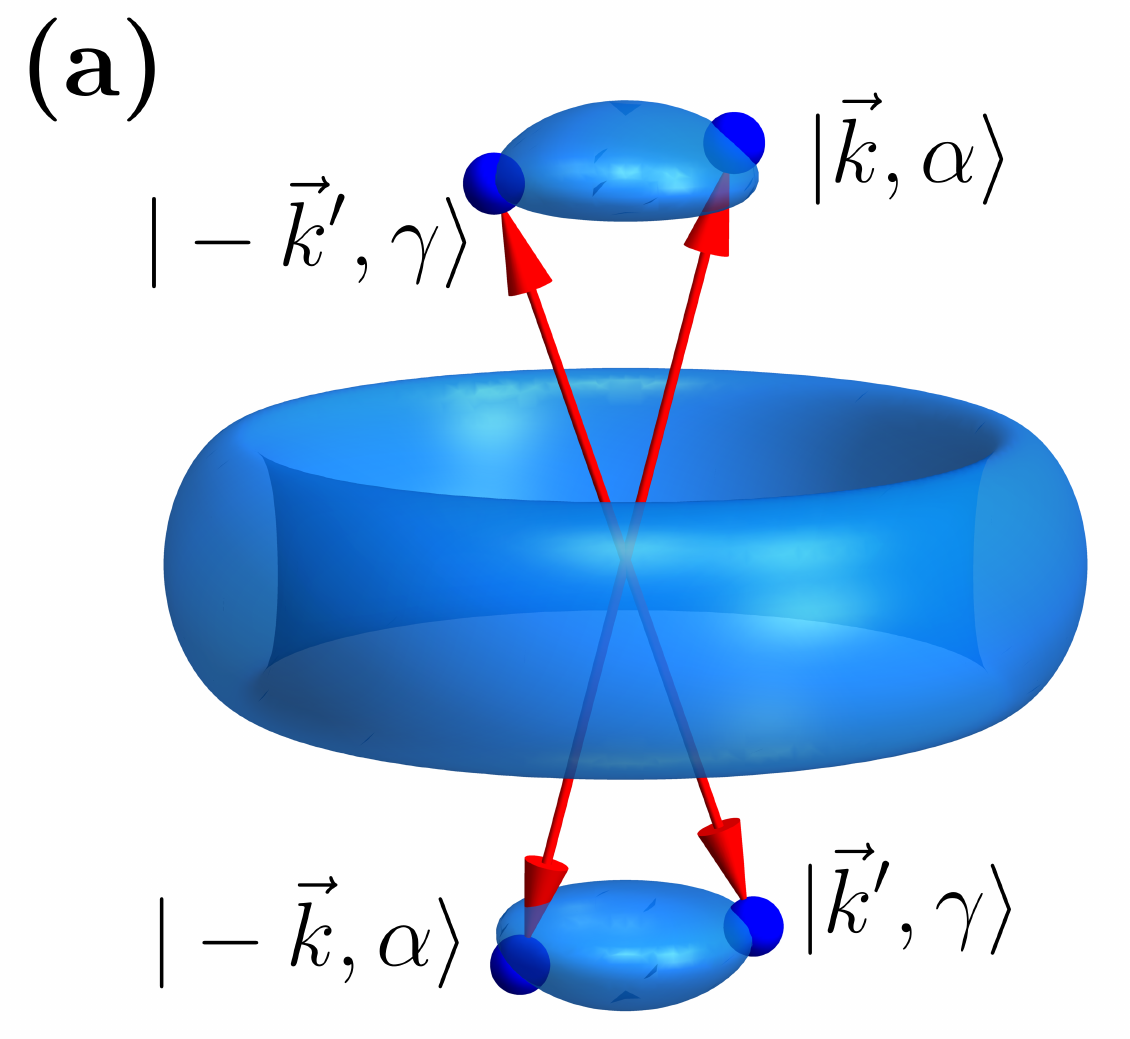}
 \includegraphics[scale=0.28]{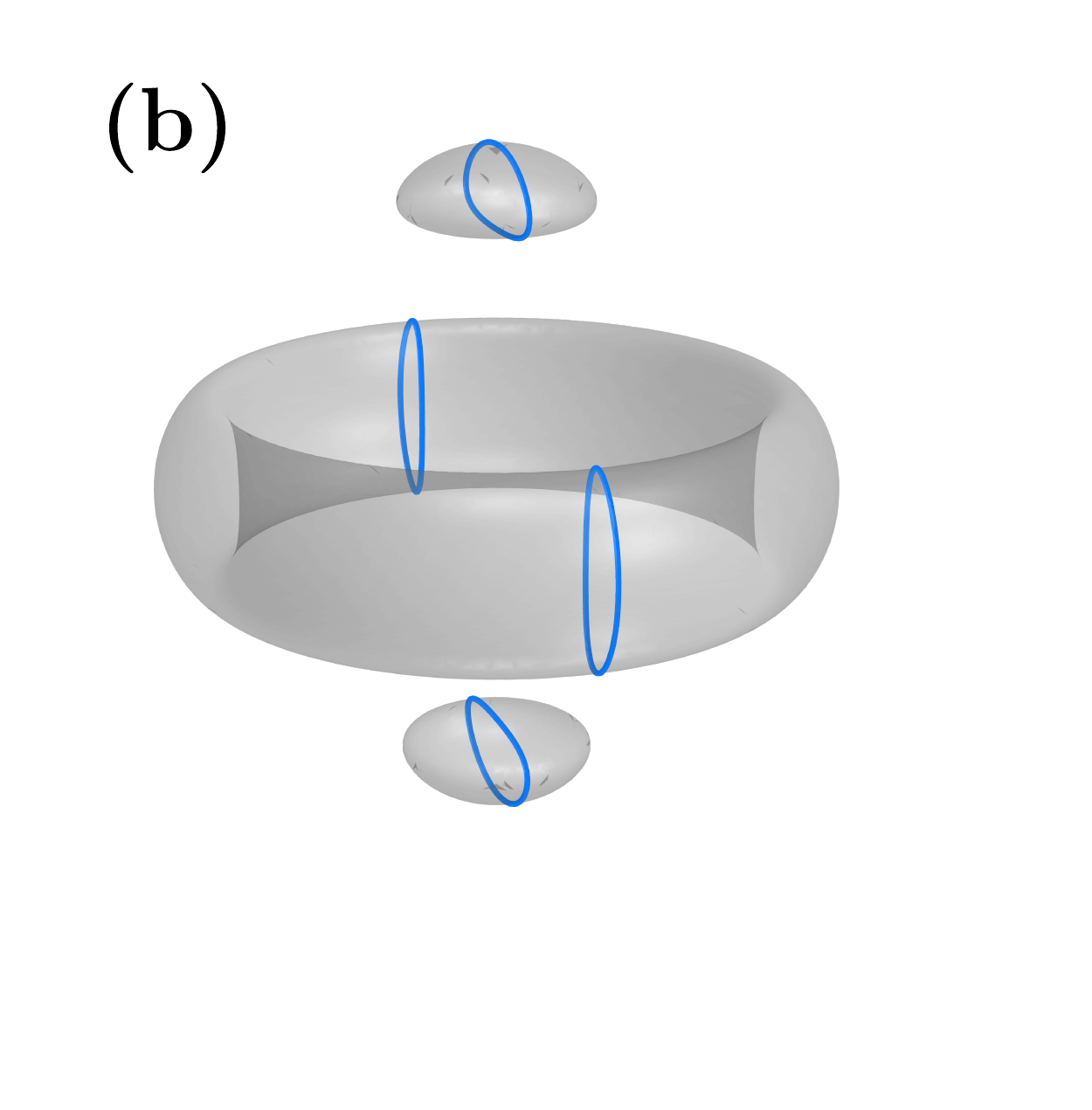}
  \includegraphics[scale=0.28]{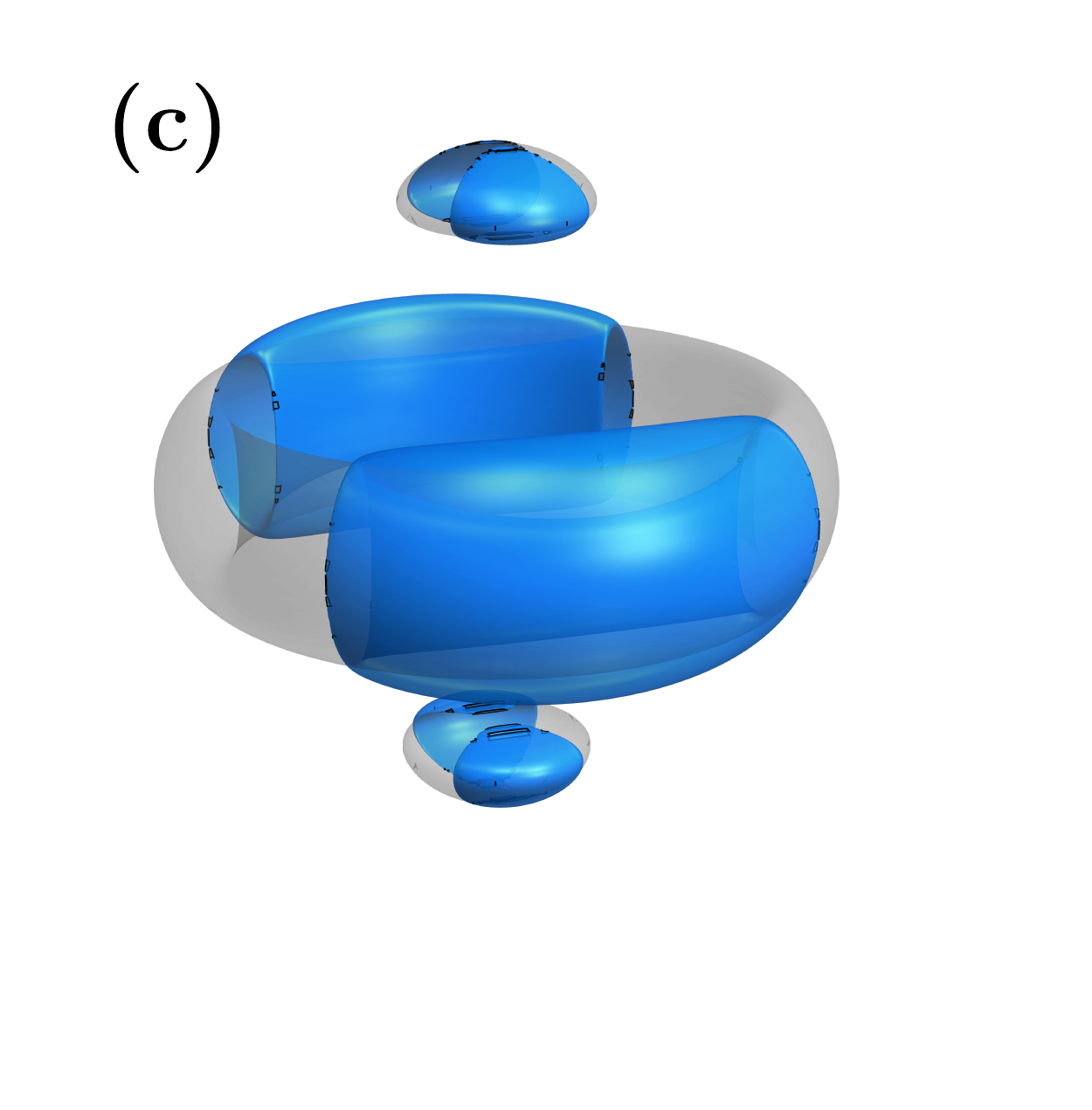}
 \caption{ (a) Example of a BG-FS (blue) in a centrosymmetric system. The inversion partners are illustrated with arrows. 
(b) Gapped BG-FS (grey) achieved by the fine-tuning condition. Line nodes (blue lines) remain. 
(c) A non-centrosymmetric BG-FS appears generically with the instability of inversion symmetry.
 }\label{F1}
\end{figure}

{\it One Bogoliubov pair problem} : 
Let us consider a generic BG-FS with inversion symmetry. Two            $\alpha$, for example angular momentum, $(| +\vec{k}, \alpha \rangle, | -\vec{k}, \alpha \rangle )$  are inversion partners.  
With  the inversion symmetry unitary operator $U_{\mathrm{Inv}}$, the single particle Hamiltonian with the superscript $(1)$ is characterized by
\begin{eqnarray}
H_{B}^{(1)} | \vec{k}, \alpha \rangle = \epsilon_k| \vec{k}, \alpha \rangle, \quad U_{\mathrm{Inv}} | +\vec{k}, \alpha \rangle =| -\vec{k}, \alpha \rangle. \nonumber
\end{eqnarray}
The inversion symmetry of the BG-FS, $[H_{B}^{(1)},U_{\mathrm{Inv}} ]=0 $, guarantees  $\epsilon_{+\vec{k}}(\alpha)= \epsilon_{-\vec{k}}(\alpha)$. 

We define one BG pair problem of the inversion partners  as a bound state quantum mechanics problem between the partners, 
\begin{eqnarray}
\Big(H_{B,1}^{(1)} + H_{B,2}^{(1)}  + V \Big) |\Psi^{(2)} \rangle = E |\Psi^{(2)} \rangle  \nonumber
\end{eqnarray}
 where an interaction between pairs, $V$, is introduced. The superscript $(2)$ is to specify a two-body problem. 
 Solving the quantum mechanical problem is standard. 
 Introducing a gap function, $\Gamma(\vec{k},\alpha,\beta) \equiv (E- \epsilon_{\vec{k}}(\alpha)- \epsilon_{-\vec{k}}(\beta)) \times \langle  \vec{k},-\vec{k}; \alpha ,\beta |\Psi^{(2)} \rangle$, we have the integral equation, 
\begin{eqnarray}
\Gamma(\vec{k}, \alpha,\beta) = \sum_{\vec{k}',\gamma,\delta} \frac{V_{\alpha \beta \gamma \delta}(\vec{k},\vec{k}')}{E-  \epsilon_{\vec{k}'}(\gamma)-\epsilon_{\vec{k}'}(\delta)} \Gamma(\vec{k}', \gamma,\delta).\ \ 
\label{gapeqn}
\end{eqnarray}
with $V_{\alpha \beta \gamma \delta}\vec{k},\vec{k}' \equiv \langle \vec{k},-\vec{k}; \alpha,\beta |V  | \vec{k}',-\vec{k}'; \gamma,\delta \rangle$.
The two particle states, $|\vec{k},-\vec{k}; \alpha, \beta \rangle $, whose quantum numbers are $(\vec{k}, \alpha)$ and $(-\vec{k}, \beta)$, is introduced.  

We are interested in a pairing between quasi-particles with the same energy, and generically it is safe to consider a case of $\alpha=\beta$ and $\gamma=\delta$. The antisymmetric property of fermions restricts a form of a gap function of the BG pair; only odd-parity functions are allowed. 
As a proof of concept, we consider a pairing potential, 
$V_{\alpha \alpha \alpha \alpha }(\vec{k},\vec{k}') =\sum_{m=0,\pm1}\lambda^{l=1}_{\alpha\alpha} w^{l=1}_{k,\alpha}w_{k',\alpha}^{l*}Y_{1,m}(\Omega_{\vec{k}})Y^{*}_{1,m}(\Omega_{\vec{k}'})$
assuming a SO(3) symmetry in energy spectrum. We omit the band index $\alpha$ below and its generalization is discussed in SM. 
Note that the structure of the integral equation is similar to the original Cooper pair problem \cite{Cooper}, and the standard self-consistent equation is obtained, 
\begin{eqnarray}
\frac{1}{|\lambda^{l=1}|} = - \sum_{\vec{k}} \frac{|w^{l=1}_k|}{E - 2 \epsilon_{\vec{k}}} \label{bound}
\end{eqnarray} 
where the summation on the right hand side is logarithmically divergent for $E >0$. For $E<0$, a bound state energy may be determined by $E = -2\Lambda e^{-2/ |\lambda^{l=1}| N_F(0) W}$ where a density of states at zero energy $N_F(0)$ of a centrosymmetric BG-FS, an averaged function over a Fermi-surface $W \equiv \langle |w^{l=1}_k|^2 \rangle_{FS}$, and a high energy cutoff $\Lambda$ are introduced. 

Few remarks are as follows. 
First, the existence of a bound state implies that a centrosymmetric BG-FS becomes unstable under an infinitesimally attractive interaction between BG quasiparticles as in the original Cooper problem. Second, the pair of BG quasiparticles is non-trivial under inversion symmetry while a Cooper pair of a Fermi-liquid is non-trivial under continuous charge conservation symmetry. Thus, our calculations indicate that a discrete symmetry is enough to induce an instability of a BG-FS.
Third, the summation of Eqn. \eqref{bound} gives a logarithmic divergence for $E\ge0$ which may be connected with the standard BCS logarithm as shown below.  Fourth,  our calculations may be generalized into a system with a lower symmetry than SO(3) and a generic pairing potential form (see SM). The former may be achieved by replacing the quantum numbers ($l,m$) with a generic representation index, and the latter may be argued by relying on the Kohn-Luttinger effect \cite{KL}. We stress that the essential part of a pair formation is the presence of a BG-FS, as in the Cooper pair problem on a Fermi liquid \cite{Cooper}.

 {\it Model Hamiltonian} : 
 We consider a model BG-FS Hamiltonian with normal and superconducting parts. 
 To be specific, we choose to follow the literatures, for example \cite{Agterberg1}, and use the model for illustrations.
 For the normal part, a Luttinger Hamiltonian in a cubic system is considered   $H_{N}=\sum_{\vec{k}}\xi_{\vec{k}}^{\dagger}H_{0}(\vec{k})\xi_{\vec{k}}$, where the explicit form is given by 
\begin{eqnarray}
H_{0}(\vec{k})&=&\left(\tilde{c}_0 \vec{k}^{2}-\epsilon_{\mathrm{F}}\right)\gamma^{0}+\sum^{3}_{a=1}\tilde{c}_1d_{a}(\vec{k})\gamma^{a} +\sum^{5}_{a=4}\tilde{c}_2 d_{a}(\vec{k})\gamma^{a}. \nonumber
\end{eqnarray}
The $4\times 4$ identity matrix  $\gamma^{0}$ is used, and $\gamma^{a}$ are five Dirac gamma matrices forming the Clifford algebra, and  a four component spinor $\xi_{\vec{k}}=(f_{\vec{k},\frac{3}{2}},f_{\vec{k},\frac{1}{2}},f_{\vec{k},-\frac{1}{2}},f_{\vec{k},-\frac{3}{2}})^{T}$ is implicitly used with fermionic annihilation  operators, $f_{\vec{k}, \alpha= \pm3/2, \pm1/2}$. 
The four parameters of the Luttinger Hamiltonian are chemical potential $\epsilon_{\mathrm{F}}$ and $\tilde{c}_0,\tilde{c}_1,\tilde{c}_2$ for particle-hole and cubic anisotropies.  
The five functions $d_{1}(\vec{k})=\sqrt{3}k_{x}k_{y}$, $d_{2}(\vec{k})=\sqrt{3}k_{y}k_{z}$, $d_{3}(\vec{k})=\sqrt{3}k_{z}k_{x}$, $d_{4}(\vec{k})=\frac{\sqrt{3}}{2}\left(k_{x}^2-k_{y}^{2}\right)$, and $d_{5}(\vec{k})=\frac{1}{2 }\left(2k_{z}^2-k_{x}^2-k_{y}^{2}\right)$ are used. 
For the superconducting part, we introduce a Nambu spinor $\chi^{T}_{\vec{k}}=(\xi^{T}_{\vec{k}},\xi_{-\vec{k}}^{\dagger})$ and  
the Hamiltonian becomes
 \begin{eqnarray}
\mathcal{H}^0_{\vec{k}}&=&\begin{pmatrix} 
H_{0}(\vec{k}) & \Delta(\vec{k}) \\
\Delta^{\dagger}(\vec{k})  & -H_{0}^{T}(-\vec{k}). 
\end{pmatrix}, \label{E2}
 \end{eqnarray}
We choose the chiral pairing channel, $\Delta(\vec{k})= \Delta_0 (\gamma^{3}+i \gamma^2)i\gamma^{12}$ with a SO(3) symmetric band structure ($c_1=c_2$) and a constant pairing $\Delta_0 \neq 0$ of the literature \cite{Agterberg1}. The contour of zero-energy states is illustrated in Fig. \ref{F1}(a).
Note that  the BG-FS Hamiltonian  enjoys the particle-hole and inversion symmetries, giving the conditions, $H_{0}(\vec{k})=H_{0}(-\vec{k})$ and $\Delta(\vec{k}) = \Delta(-\vec{k})=-\Delta^{T}(-\vec{k})$. But, the time-reversal symmetry is explicitly broken as shown in the form of $\Delta(\vec{k})$. The symmetries constrain the eight band spectrums of $\mathcal{H}^0_{\vec{k}}$. Below, whenever necessary, we take $\mathcal{H}^0_{\vec{k}}$ as our microscopic Hamiltonian, and yet we stress that our discussions about interaction effects are generic.

{\it Interaction effects :}  
Our strategy to investigate interaction effects of a BG-FS is as follows. First, we construct an effective two-band model of a BG-FS. Second, we employ standard mean-field calculation and renormalization group analysis. Then, we investigate implications of our results in terms of a phenomenological theory. 

Let us construct an effective two-band model. It is crucial to notice that the zero energy states are doubly degenerate at each momentum because of the particle-hole and inversion symmetries. 
Therefore, a two-band model is inevitable to capture low energy excitations, and we introduce an effective low energy Hamiltonian of a centrosymmetric BG-FS, $\mathcal{H}_{\vec{k}, 0}^{\mathrm{eff}} = E_0(\vec{k}) \tau^z$, with a two component spinor $\Psi_{\vec{k}}$. 
The Pauli-matrices ($\tau^{x,y,z}$) are to describe the two-band space, where the particle-hole and inversion symmetries may have the forms, $U_c = \tau^x \mathcal{K}$ and  $U_{\mathrm{Inv}} = \tau^0\equiv I_{2\times2}$, respectively. $\mathcal{K}$ is the standard complex conjugation operator.  One may obtain the effective two-band Hamiltonian and symmetry properties by projecting the microscopic eight-band model, $\mathcal{H}^0_{\vec{k}}$ onto the two-band space (see SM). 
It is useful to study how an order parameter of inversion symmetry, $\phi$, is coupled to operators in our effective model. 
The Hermitian properties constrain the coupling significantly, 
\begin{eqnarray}
\mathcal{H}_{\vec{k},0}^{\mathrm{eff}} \rightarrow \mathcal{H}_{\vec{k}}^{\mathrm{eff}} (\phi) \equiv\mathcal{H}_{\vec{k},0}^{\mathrm{eff}}- \phi  \sum_{\mu=0,x,y,z}  \rho_{\mu}(\vec{k}) \tau^{\mu}, 
\end{eqnarray}
in the two-band space.  The inversion symmetry imposes the odd parity conditions, ($\rho_{\mu}(\vec{k})=-\rho_{\mu}(-\vec{k})$). Furthermore, $\tau^z$ is odd under the particle-hole symmetry transformation. Thus, one of the odd parity functions vanishes, $\rho_z(\vec{k})=0$. The energy spectrum of the non-interacting Hamiltonian is $ \big( \sqrt{E_0(\vec{k})^2 + \phi^2   \rho_x(\vec{k})^2 } - \phi \rho_0(\vec{k}) \big)$ at a given $\vec{k}$.
Next, we incorporate interactions and consider a generic short-range interaction of a centrosymmetric BG-FS, whose form may be expressed as 
\begin{eqnarray}
H_{\mathrm{tot}}=  H^{\mathrm{eff}}_0  - \frac{1}{2}\sum_{\mu,\nu;\vec{k},\vec{k}'}  g_{\mu\nu} V_{\mu \nu}(\vec{k},\vec{k}') (\Psi^{\dagger}_{\vec{k}} \tau^{\mu} \Psi_{\vec{k}}) ( \Psi^{\dagger}_{\vec{k}'} \tau^{\nu} \Psi_{\vec{k}'} ) \nonumber
\end{eqnarray}
with a two component spinor $\Psi_{\vec{k}}$ and $H^{\mathrm{eff}}_0 = \sum_{\vec{k}} \Psi^{\dagger}_{\vec{k}} \big( \mathcal{H}_{\vec{k},0}^{\mathrm{eff}} \big)\Psi_{\vec{k}}$. For simplicity, we consider a separable interaction $V_{\mu \nu}(\vec{k},\vec{k}')= \rho_{\mu}(\vec{k}) \rho_{\nu}(\vec{k}')$ and later we argue its generalization.
 The particle-hole symmetry imposes the conditions $(g_{z0}=g_{zx}=g_{zy}=0)$. 
On the other hand, the mixing terms ($g_{x0}, g_{y0}, g_{xy}$) are generically non-zero unless an extra constraint is imposed. 
 
We perform the standard mean-field analysis for the effective two-band model, $H_{\textrm{tot}}$, with an ansatz, $  \langle   \Psi^{\dagger}_{\vec{k}}  \tau^{\mu}\Psi_{\vec{k}}\rangle_{\mathrm{MF}} \equiv \phi \, c^{\mu}_{\vec{k}}$, and find the mean-field Hamiltonian, 
\begin{eqnarray}
H_{\mathrm{MF}}&=&H^{\mathrm{eff}}_{0}-\phi\sum_{\mu, \vec{k}}d_{\mu}\rho_{\mu}(\vec{k}) \Psi^{\dagger}_{\vec{k}}  \tau^{\mu}\Psi_{\vec{k}} +\frac{\phi^{2}}{2}\sum_{\mu,\nu}d_{\mu} g_{\mu\nu}^{-1}d_{\nu}, \nonumber 
\end{eqnarray}
 with $d_{\mu} \equiv \sum_{\nu ,\vec{k}}g_{\mu\nu} \rho_{\nu}(\vec{k}) c^{\nu}_{\vec{k}} $. 
The inverse matrix of $g_{\mu\nu}$ ($\sum_{\nu}g_{\mu \nu}g_{\nu \sigma}^{-1}= \delta_{\mu \sigma}$) is introduced whose determinant is generically non-zero. 

To be specific, we consider a Hamiltonian with three coupling constants ($g_{xx}, g_{x0}, g_{00}$) for simplicity. 
The Hamiltonian may be diagonalized by introducing a unitary transformation, 
\begin{eqnarray}
H_{\mathrm{MF}} &=& \sum_{\vec{k},\alpha=\pm} E(\vec{k};\phi) \,\gamma^{\dagger}_{\alpha}(\vec{k})  \gamma_{\alpha}  (\vec{k}) +\mathcal{E}_0(\phi),
\end{eqnarray}
with 
\begin{eqnarray}
E(\vec{k};\phi) &\equiv& \sqrt{E_0(\vec{k})^2 + \phi^2 d_{x}^{\:2} \rho_x(\vec{k})^2 }-\phi d_0\rho_0(\vec{k}), \nonumber 
\end{eqnarray}
and
\begin{eqnarray}
\mathcal{E}_0(\phi) &\equiv& - \sum_{\vec{k}} \sqrt{E_0(\vec{k})^2 + \phi^2 d_{x}^{\:2} \rho_x(\vec{k})^2 } + \frac{\phi^{2}}{2}\sum_{\mu,\nu}d_{\mu} g_{\mu\nu}^{-1}d_{\nu}.\nonumber 
\end{eqnarray}
The unitary transformation of $\Psi_{\vec{k}}$ determines the creation/annihilation operators ($\gamma_{\alpha}(\vec{k}), \gamma^{\dagger}_{\alpha}(\vec{k})$) with $\alpha =\pm$.

The ground state energy is 
$E_{G}^{\mathrm{MF}}[\phi]  = \mathcal{E}_0(\phi) +  2 \sum_{\vec{k} \in M^-} E(\vec{k};\phi)$, where the negative energy manifold $M^-$ is specified by the condition $E(\vec{k};\phi)<0$.
A manifold of zero energy excitations ($E(\vec{k};\phi)=0$) is obtained by the condition, 
\begin{eqnarray}
E_0(\vec{k})^2 + \phi^2 d_{x}^{\:2} \rho_x(\vec{k})^2 = \phi^2 d_0^2\rho_0(\vec{k})^2,
\end{eqnarray}
which gives a Fermi-surface generically. 
 In Fig. \ref{F2}, a mean-field phase diagram is obtained by minimizing the mean-field free energy, $F_{\mathrm{MF}} = -T \log ({\rm Tr} (e^{- H_{\mathrm{MF}}/T} ))$. 


Main results of our mean field calculations may be summarized as follows. 
First, a centrosymmetric BG-FS is absent at zero temperature $T=0$, and thus the inversion symmetry breaking is instability of a centrosymmetric BG-FS. The ground state energy difference is defined as $\Delta E_{G}^{\mathrm{MF}}[\phi] \equiv E_{G}^{\mathrm{MF}}[\phi] - E_{G}^{\mathrm{MF}}[0]$ whose explicit form is 
\begin{eqnarray}
\Delta E_{G}^{\mathrm{MF}}[\phi] = \sum_{\vec{k}}\Big[ |E_0(\vec{k})|-\sqrt{E_0(\vec{k})^2 + \phi^2 d_{x}^{\:2} \rho_x(\vec{k})^2 } \Big] +\cdots. \nonumber
\end{eqnarray}
The first term gives the BCS type instability, as manifested in the logarithmic divergence, 
\begin{eqnarray}
\frac{\partial \Delta E_{G}^{\mathrm{MF}}[\phi]}{\partial \phi^2}|_{\phi=0} = -\frac{d_{x}^{\:2} }{2} \sum_{\vec{k}} \frac{ \rho_x(\vec{k})^2 }{|E_0(\vec{k})|} \propto -\log(\frac{\Lambda}{\mu}) \nonumber
\end{eqnarray}
where the high/low energy cutoffs ($\Lambda/\mu$) are introduced. Thus, the inversion symmetry must be broken at $T=0$.
Second, a centrosymmetric BG-FS survives at non-zero temperature whose regime diminishes at lower temperatures as shown in Fig. \ref{F2}. 
Third, the original Fermi-surface is transformed by the inversion symmetry breaking, and a Fermi-surface of BG quasiparticles survives unless fine-tuned, for example $\rho_0(\vec{k})=0$, is imposed (see SM for detailed discussion about the fine-tuning condition). In Fig. \ref{F1} (b), one example of a fine-tuned case ($\rho_{x}(\vec{k})=\rho_{0}(\vec{k})=k_{x}/|\vec{k}|$) is illustrated. Generically, the two functions ($\rho_0(\vec{k}), \rho_x(\vec{k})$) are independent even for a separable interaction. We stress that the survival of a Fermi-surface after the instability of inversion symmetry is drastically different from the standard BCS superconductivity, manifested by the presence of the $\tau^0$ channel.  In Fig. \ref{F1} (c), one example of a non-centrosymmetric BG-FS is illustrated, which clearly shows the presence of a Fermi-surface ($\rho_{x}(\vec{k})=\rho_{0}(\vec{k})=k_{x}/|\vec{k}|$).  

%

\begin{figure}[tb]
\centering
\includegraphics[scale=0.45]{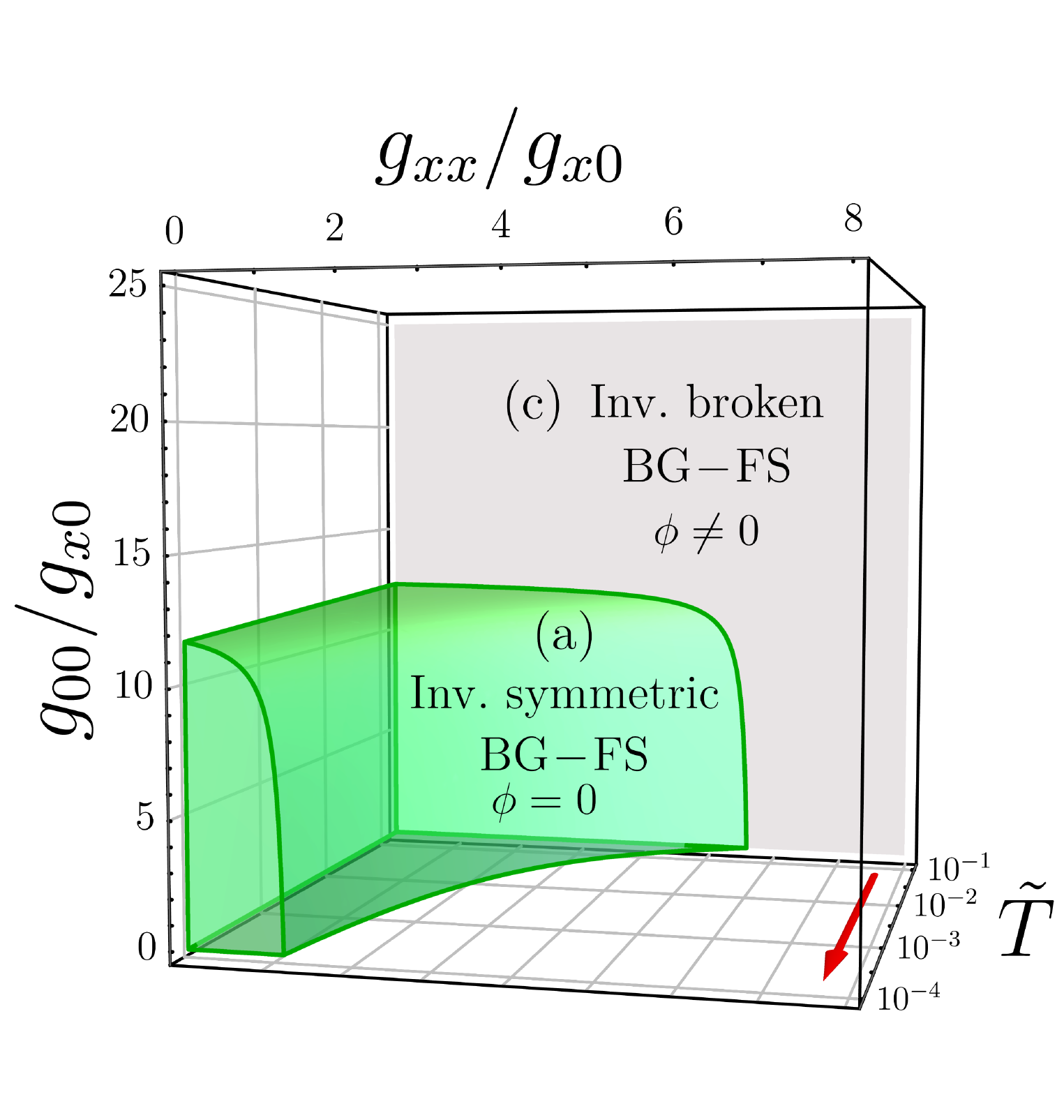}  \ \  \vspace{-10pt}
\caption{Schematic mean-field phase diagrams at different temperatures. Parameters of the separable interactions are $\rho_0(\vec{k}) =k_y / |\vec{k}|, \rho_x(\vec{k}) =k_x / |\vec{k}|$. The relative coupling constants, $ (g_{xx}/g_{xo}) $, $(g_{00}/g_{xo})$, and dimensionless temperature, $\tilde{T}\equiv T/\Lambda$ with a UV-cutoff scale are introduced for each axes. A centrosymmetric BG-FS is stable for weak coupling regions (green) at each temperature and becomes unstable for strong coupling regimes where its inversion symmetry is broken ($\phi\neq0$). It is clear that the regions of a BG-FS shrink as lowering temperature (red arrow) and eventually vanish in the limit of $T\rightarrow0$.  
} \label{F2}
\end{figure}
 
To go beyond the mean-field analysis, we perform the standard renormalization group analysis. For simplicity, we illustrate the case with the three coupling constants, and the generic cases with six coupling constants are discussed in SM. Introducing dimensionless coupling constants, $\tilde{g}_{\mu\nu}$ which are averaged quantities over a Fermi-surface weighted by $\rho_{\mu}(\vec{k})$, we find   
\begin{eqnarray}
\frac{d \tilde{g}_{xx}}{dl }&=& \tilde{g}_{xx}^2, \quad \frac{d \tilde{g}_{x0}}{dl }=\tilde{g}_{x0}\tilde{g}_{xx}, \quad \frac{d \tilde{g}_{00}}{dl }= \tilde{g}_{x0}^2, 
\end{eqnarray}
with the scale variable $l$ of renormalization group analysis. The long wavelength limit is $l \rightarrow \infty$. In the $\tilde{g}_{xx}$ channel, the BCS type logarithmic dependence  manifests.  It is obvious that the first two equations have the positive eigenvalues, and the right-hand-side of the third one is always positive. Thus, the original BG-FS is unstable at $T=0$ for attractive bare interactions, which is consistent with the mean-field results.  
 
{\it Ginzburg-Landau Theory} : 
 The above instability calculations indicate that the inversion symmetry order parameter should be included in a phenomenological Ginzburg-Landau theory of BG-FSs from the beginning. The Ginzburg-Landau functional is
 \begin{eqnarray}
 \mathcal{F} [\Delta, \phi] = r_{\Delta} {\rm Tr} \big[\Delta^{\dagger} \Delta  \big] +r_{\phi} \phi^2 + \cdots,
 \end{eqnarray}
 which can be obtained by integrating out fermions at a non-zero temperature. A BG-FS may be considered by the condition $r_{\Delta} < 0$, and our instability calculation indicates $r_{\phi}=r_{\phi}^0 -  \langle \mathcal{O} \rangle_{\mathrm{FS}} \log(\frac{\Lambda}{T}) $ with a positive-definite quantity averaged over a BG-FS,  $\langle \mathcal{O}\rangle_{\mathrm{FS}}\propto \langle \rho_{x}^{2} \rangle_{\mathrm{FS}}$. 

 As usual, the sign of higher order terms may determine natures of transitions, continuous or discontinuous, and first-order transitions to other symmetry broken phases are possible.
 Note that the sign of the interaction term between the two order parameters, $|\Delta|^2 \phi^2$, determines whether the order parameters compete or cooperate.  
 
 Let us consider a schematic phase diagram of the Ginzburg-Landau functional. 
  Adjusting the parameters ($r_{\Delta}, r_\phi$), we may set $\mathrm{O}=(0,0)$, the multi-critical point.  Possible four phases are  
 \begin{itemize}
 \item (A) ($r_{\Delta}>0, r_{\phi}>0 $) : centrosymmetric metal,
  \item (B) ($r_{\Delta}<0, r_{\phi}>0 $) : centrosymmetric BG-FS,
   \item (C) ($r_{\Delta}>0, r_{\phi}<0 $) : polar metal,
 \item (D) ($r_{\Delta}<0, r_{\phi}<0 $) :  non-centrosymmetric SC.
 \end{itemize}
 Note that an intermediate phase between (A) and (B) may be present. For example, a time reversal symmetric superconductor may appear if (A) is a time reversal symmetric metal. 
Our instability calculations indicate that the phase (D) always appear at low temperature. 
In (D), the inversion partners of BG quasi-particles have different energy.  As discussed above, a Fermi-surface generically survives in a non-centrosymmetric BG-FS similar to the ones in literature \cite{Agterberg_,Agterberg2, Gibaik2,Fernandes}. 
Furthermore, the Ginzburg-Landau theory indicates that a phase transition from (A) to (D) generically happens with two step transitions unless it is fine-tuned to go through $\mathrm{O}$. 

{\it Discussion and Conclusion} : 
Based on our instability results, we propose enhanced fluctuations of an inversion order parameter is a key property of a  centrosymmetric BG-FS. This is analogous to the fact that a Fermi liquid is always susceptible to a superconducting instability, as shown by the seminal work by Kohn and Luttinger \cite{KL}. 
In Fig. \ref{F3}, we illustrate a schematic phase diagram with a tuning parameter of quantum fluctuations of an inversion order parameter. 
Our results indicate that a weakly interacting centrosymmetric BG-FS is unstable, and the phase \textbf{X} is \textit{absent} (See Fig. \ref{F3} (b)). 
On the other hand, it is an interesting question whether strongly interacting BG quasi-particles stabilize a centrosymmetric BG-FS because our above calculations are based on the assumption of well-defined Bougoliubov quasi-particles. 
The recent work of a pairing instability in a non-Fermi liquid \cite{Metlitski} suggests that a stable BG-FS may be possible down to zero temperature if its excitations lose quasi-particle natures. 

Enhanced fluctuations of an inversion order parameter may be captured by inversion susceptibility. An external field of the order parameter is required to measure the susceptibility. Motivated by recent advances in flexoelectricity, we note that a strain gradient on a sample breaks inversion symmetry and plays a role of an external field of an inversion order parameter. Moreover, it is well known that second harmonic generation (SHG) experiment is a probe to identify an inversion order parameter  \cite{Boyd}. 
In other words, SHG provides information of the onset of an inversion order parameter, for example, $\phi \sim (T_c - T)^{\beta}$, with the critical temperature of inversion symmetry breaking $T_c$. 
Combining the two methods, we propose a second harmonic generation experiment with a strain gradient to measure inversion susceptibility and expect to obtain information of the susceptibility, $\chi_{\phi} \sim |T_c-T|^{-\gamma}$. Note that the susceptibility has a non-trivial signatures even at higher temperatures, $T >T_c$ in sharp contrast to the absence of an order parameter at higher temperatures. We believe the SHG with a strain gradient may be applied in both superconducting and normal states with enhanced inversion fluctuations since inversion symmetry acts in the same way. It is desired to test the experiment in the candidate heavy fermion materials including URu$_2$Si$_2$ and UBe$_{13}$. 
Recently, FeSe is also proposed to be a candidate system of a BG-FS \cite{Hirschfeld}, and we believe that inversion order parameter fluctuations may be enhanced in FeSe.

\begin{figure}[tb]
\centering\vspace{10pt}
\includegraphics[scale=0.322]{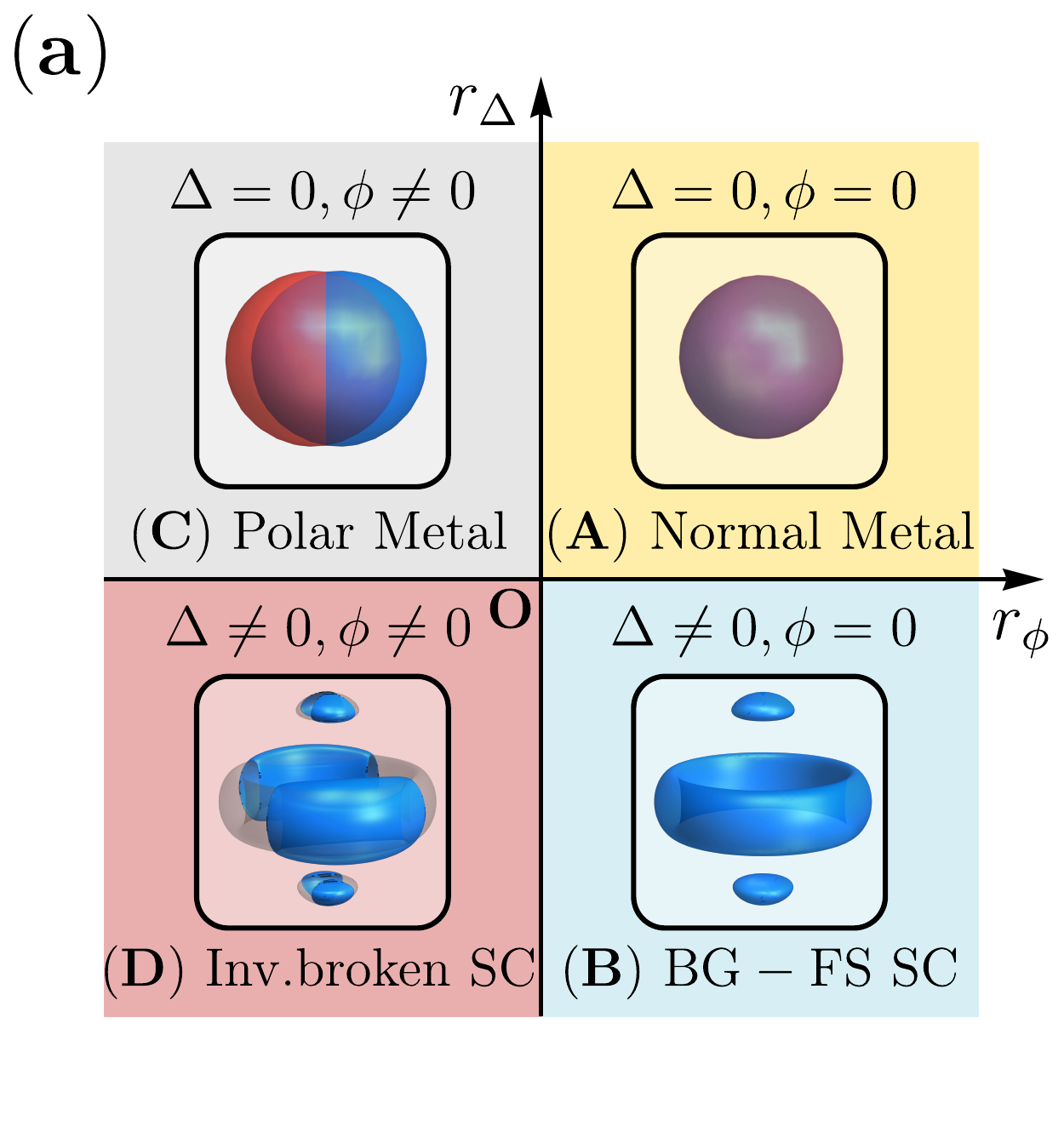} 
\includegraphics[scale=0.348]{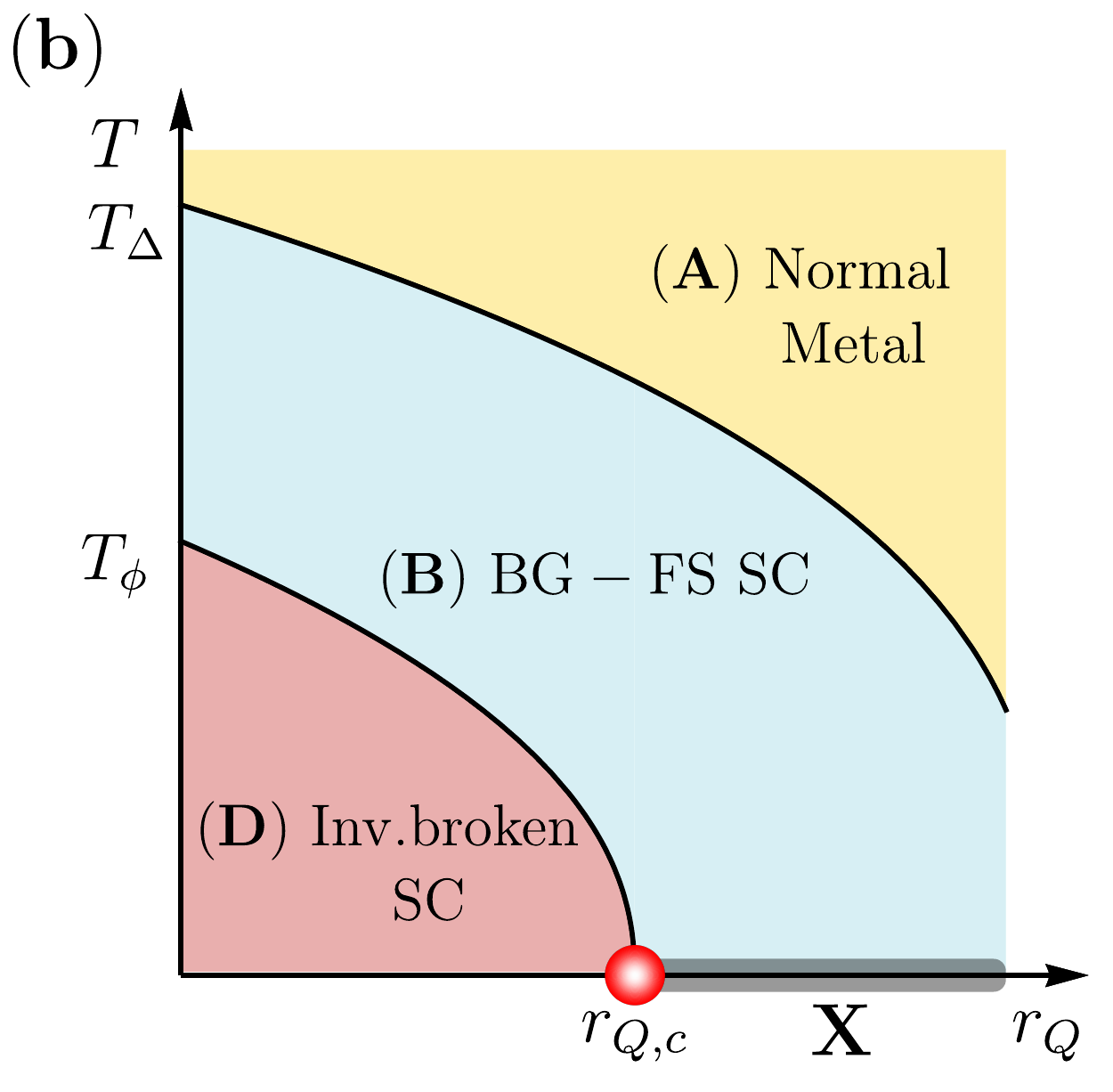}\vspace{-10pt}
 \caption{ (a) Generic phase diagram of the four phases. The phase (B) becomes unstable at low temperatures. (b) schematic phase diagram with the two parameters,  a quantum fluctuation parameter $r_Q$ and temperature $T$. The phase \textbf{X} is BG quasiparticle excitation s at zero temperature.  
 Our results indicate that the phase \textbf{X} is \textit{absent} if BG quasi-particles are well-defined on a BG-FS.
 } \label{F3}
\end{figure}

In conclusion, we investigate interaction effects of a centrosymmetric BG-FS and find its instability in the inversion symmetry channel. Condensation of BG pairs induces the instability, similar to the BCS instability of Fermi liquids where Cooper pairs condense  and break $U(1)$ symmetry.
On the other hand, in contrast to the standard BCS superconductivity \cite{Volkov}, a Fermi-surface generically survives unless fine-tuned.  
The instability enforces a phenomenological Ginzburg-Landau functional to include an inversion order parameter  from the beginning. 
Future works including disorder effects and strong quantum fluctuations are highly desired, and microscopic calculations of SHG with a strain gradient would be also useful.

\textit{Acknowledgement} :
 We thank D. Agterberg, S. E. Han, J. S. Kim, D. Lee, and T. Shibauchi for discussions. We are particularly grateful to D. Agterberg for invaluable comments and K. Hwang for critical aid for mean field calculations.
This work was supported by the POSCO Science Fellowship of POSCO TJ Park Foundation and NRF of Korea under Grant NRF-2017R1C1B2009176 and NRF-2019M3E4A1080411.

\onecolumngrid
\clearpage
\begin{center}
\textbf{\large Supplemental Material for ``Instability of $j= 3/2$ Bogoliubov Fermi-surfaces''}
\end{center} 
\begin{center} \vspace{-5pt}
{Hanbit Oh and Eun-Gook Moon$^{\textcolor{red}{*}}$}\\
\emph{Department of Physics, Korea Advanced Institute of Science and Technology, Daejeon 305-701, Korea}
\end{center}
\setcounter{equation}{0}
\setcounter{figure}{0}
\setcounter{table}{0}
\setcounter{page}{1}
\setcounter{section}{0}
\setcounter{subsection}{0}

\maketitle 
\makeatletter
\renewcommand{\thesection}{\arabic{section}}
\renewcommand{\thesubsection}{\thesection.\arabic{subsection}}
\renewcommand{\thesubsubsection}{\thesubsection.\arabic{subsubsection}}
\renewcommand{\theequation}{S\arabic{equation}}
\renewcommand{\thefigure}{S\arabic{figure}}
\renewcommand{\thetable}{S\arabic{table}}

\section{ONE BOGOLIUBOV PAIR PROBLEM} \label{SS1}
Let us consider two states which are inversion partners ($| +\vec{k}, \alpha \rangle, | -\vec{k}, \alpha \rangle$) in a centrosymmetric Bogoliubov Fermi-surface (BG-FS). The two eigenstates of a single-particle Hamiltonian are related by an inversion symmetry unitary operator $U_{\mathrm{Inv}}$,
\begin{eqnarray} U_{\mathrm{Inv}} | +\vec{k}, \alpha \rangle =| -\vec{k}, \alpha \rangle,  \quad H_{\mathrm{B}}^{(1)} | \vec{k}, \alpha \rangle = \epsilon_k| \vec{k}, \alpha \rangle. \nonumber
\end{eqnarray}
The superscript $(1)$ specifies an one-particle Hamiltonian, and inversion symmetry indicates $\epsilon_{\vec{k}}(\alpha)= \epsilon_{-\vec{k}}(\alpha)$. The parameter $\alpha$ is for an additional quantum number such as a spin degree of freedom.  

 We define one Bogoliubov pair problem as a quantum mechanics bound state problem of a two-particle state $|\Psi^{(2)} \rangle$, which is specified by the superscript $(2)$. The Schr$\mathrm{\ddot o}$dinger equation of the two particles is 
\begin{eqnarray}
\Big(H_{\mathrm{B},1}^{(1)} + H_{\mathrm{B},2}^{(1)}  + V \Big) |\Psi^{(2)} \rangle = E |\Psi^{(2)} \rangle,  \nonumber
\end{eqnarray}
 and the gap integral equation is
\begin{eqnarray}
\Gamma(\vec{k}, \alpha,\beta) = \sum_{\vec{k}',\gamma,\delta} \frac{V_{\alpha \beta \gamma \delta}(\vec{k},\vec{k}')}{E- \epsilon_{\vec{k}'}(\gamma)-\epsilon_{\vec{k'}}(\delta)} \Gamma(\vec{k}', \gamma,\delta),
\label{se2}
\end{eqnarray}
with  the gap function, 
\begin{eqnarray}
 \Gamma(\vec{k},\alpha,\beta) &\equiv& (E- \epsilon_{\vec{k}}(\alpha)-\epsilon_{-\vec{k}}(\beta) )\  \langle  \vec{k},-\vec{k}; \alpha ,\beta |\Psi^{(2)} \rangle.
 \end{eqnarray}
A pairing matrix element between the two-particle states ($|\vec{k},-\vec{k}; \alpha, \beta \rangle $, $|\vec{k}',-\vec{k}'; \gamma, \delta \rangle $) is introduced, 
$V_{\alpha \beta \gamma \delta}(\vec{k},\vec{k}')\equiv \langle \vec{k},-\vec{k}; \alpha,\beta |V  | \vec{k}',-\vec{k}; \gamma,\delta \rangle$. 
The structure of the integral equation is similar to that of an original Cooper pair problem.

Considering a system with SO(3) symmetry, we may use isotropic kinetic energy, $\epsilon_{\vec{k}}(\alpha)\equiv \epsilon_{k}(\alpha)$ with $k\equiv |\vec{k}|$, and the wave functions and interaction potentials may be decomposed by the angular momentum quantum numbers ($l,m$) of SO(3) group. 
Since a BG-FS beaks time-reversal symmetry, a bound state between inversion partners is considered in contrast to the original Cooper pair problem.
Following the Cooper's analysis, we focus on the $ \Gamma(\vec{k},\alpha,\beta) \propto \delta_{\alpha\beta}$ and $V_{\alpha \beta \gamma \delta}(\vec{k},\vec{k}')\propto \delta_{\alpha\beta}\delta_{\gamma\delta}$ case, which may be justified by the same energy condition. 
This leads to 
\begin{eqnarray}
\langle  \vec{k},-\vec{k}; \alpha ,\alpha |\Psi^{(2)} \rangle  &=& \sum_{l: {\rm odd},m}a_{k,\alpha}^{l}Y_{l,m}(\Omega_{\vec{k}}),
\end{eqnarray}\vspace{-10pt}
\begin{eqnarray}
\langle \vec{k},-\vec{k}; \alpha,\alpha |V  | \vec{k}',-\vec{k}'; \gamma,\gamma \rangle&=&-\sum_{l: {\rm odd},m}|\lambda^{l}_{\alpha\gamma}|w^{l}_{k,\alpha}w_{k',\gamma}^{l*}Y_{l,m}(\Omega_{\vec{k}})Y^{*}_{l,m}(\Omega_{\vec{k}'}).
\end{eqnarray}
 We take an attractive factorizable potential with the interaction strength $|\lambda^{l}_{\alpha\gamma}|$ and assume that the intra-coupling constants are much larger than the inter-coupling constants, $|\lambda^{l}_{\alpha\alpha}|\gg |\lambda^{l}_{\alpha\neq\gamma}|$, for simplicity. The even $l$ channels are forbidden due to the antisymmetric nature of a two-fermion wave function, 
\begin{eqnarray}
\langle  \vec{k},-\vec{k}; \alpha ,\alpha |\Psi^{(2)} \rangle=-\langle  -\vec{k},\vec{k}; \alpha,\alpha  |\Psi^{(2)} \rangle,
\end{eqnarray}
and the parity of spherical harmonics, $Y_{l,m}(\Omega_{\vec{k}})=(-1)^lY_{l,m}(\Omega_{-\vec{k}})$.

The self-consistent equation with the quantum number $(l,\alpha)$ is
\begin{eqnarray}
\frac{1}{|\lambda^{l}_{\alpha\alpha}|}&=&-\sum_{k}\frac{|w_{k,\alpha}^{l}|^{2}}{E-2\epsilon_{k}(\alpha)}=-N_{\alpha}(0)\langle|w_{k,\alpha}^{l}|^{2}\rangle_{\mathrm{FS}}\int^{\Lambda}_{0} \frac{d\epsilon}{E-2\epsilon}.
\end{eqnarray}
A non-zero density of states (DOS) of a BG-FS at zero energy, $N_{\alpha}(0) $, and the averaged quantity over a BG-FS, $ W^{l}_{\alpha}\equiv \langle|w_{k,\alpha}^{l}|^{2} \rangle_{\mathrm{FS}}$, are introduced. The energy integration has  a high energy cut-off, $\Lambda$, and  the integral equation has one solution for each $(l,\alpha)$,
\begin{eqnarray}
E^{l}_{\alpha}&=&-2\Lambda \exp\left(-\frac{2}{|\lambda^{l}_{\alpha\alpha}|N_{\alpha}(0)W^{l}_{\alpha}} \right).
\end{eqnarray} 
We use the weak coupling condition, $|\lambda^{l}_{\alpha\alpha}| N_{\alpha}(0) W^{l}_{\alpha} \ll 1$. Note that the solution is for a bound state, manifested by the negative sign of the solution.  

One may generalize our results to a generic case with a discrete point group symmetry $G$. Then, the wave functions and interactions may be decomposed into  basis functions of an irreducible representation $R$ of group $G$ instead of spherical harmonics $Y_{l}^{m}(\Omega_{\vec{k}})$ for the SO(3) group. 
%

\section{ FERMIONIC HAMILTONIAN OF NORMAL STATE}
We consider a system with cubic and time-reversal symmetries which may realize a quadratic band touching energy spectrum in three spatial dimensions. The low energy Hamiltonian, so-called Luttinger Hamiltonian, is 
\begin{eqnarray}
H_{0}(\vec{k}) & = &(\hat{c}_{0}k^{2}-\epsilon_{\mathrm{F}})\gamma^{0}+\sum_{a=1}^{5}\hat{c}_{a}d_{a}(\vec{k})\gamma^{a}.  \label{se12}
\end{eqnarray}
The quadratic functions $d_a({\vec{ k}})$ and four dimensional Gamma matrices ($\gamma^{a}$) are, 
 \begin{eqnarray}
d_{1}(\vec{k})=\sqrt{3}k_{x}k_{y},\  d_{2}(\vec{k})=\sqrt{3}k_{y}k_{z},\ d_{3}(\vec{k})=\sqrt{3}k_{z}k_{x},\ d_{4}(\vec{k})=\frac{\sqrt{3}}{2}\left(k_{x}^2-k_{y}^{2}\right),\  d_{5}(\vec{k})=\frac{1}{2 }\left(2k_{z}^2-k_{x}^2-k_{y}^{2}\right), \nonumber
\end{eqnarray}
and
\begin{eqnarray}
\gamma^{1}=\sigma^{y}\otimes \sigma^{0},\ \ \gamma^{2}=\sigma^{z}\otimes \sigma^{y},\ \  \gamma^{3}=\sigma^{z}\otimes \sigma^{x},\ \ \gamma^{4}=\sigma^{x}\otimes \sigma^{0},\ 
\gamma^{5}=\sigma^{z}\otimes \sigma^{z}. \nonumber
\end{eqnarray} 
where a Clifford algebra $\{\gamma^{a},\gamma^{b}\}=2\delta_{ab}$, and $\gamma^{ab}\equiv\frac{1}{2i}[\gamma^{a},\gamma^{b}]$ are introduced \cite{Murakami}. It may be expressed in terms of $j=\frac{3}{2}$ angular momentum operators, 
\begin{eqnarray}
J_{1}= \frac{\sqrt{3}}{2}\gamma^{25}+\frac{1}{2}(\gamma^{13}+\gamma^{24}),\ \ J_{2}= -\frac{\sqrt{3}}{2}\gamma^{35}-\frac{1}{2}(\gamma^{12}-\gamma^{35}),\ \ J_{3}= -\gamma^{14}-\frac{1}{2}\gamma^{23}. \nonumber
\end{eqnarray}
The cubic symmetry allows that Luttinger Hamiltonian is parameterized by four parameters, chemical potential $\epsilon_{\mathrm{F}}$, and  $(\tilde{c}_0=\hat{c}_{0}, \tilde{c}_1= \hat{c}_{1,2,3},\tilde{c}_2=\hat{c}_{4,5})$.
The doubly degenerate energy eigenvalues are,
\begin{eqnarray}
E_{0,\nu}(\vec{k}) =(\tilde{c}_{0}k^{2}-\mu) +\nu \sqrt{\sum_{a} \tilde{c}_a^2 d_a(\vec{k})^2},  \ \ \ \mathrm{for\ } \nu=\pm1. \label{se9}
\end{eqnarray}


\section{Bogoliubov Fermi-surfaces with $J=3/2$ systems}\label{s3}

We consider model Hamiltonians with different numbers of bands, named \textit{Eight-}, \textit{four-}, and \textit{two-band models}. Since we are interested in superconductivities, the sizes of the Bogoliubov-de Gennes (BdG) Hamiltonians are doubled from the number of physical energy.  
For example, two-band Hamiltonian is described by a two component Nambu spinor, which is composed of a quasiparticle and quasihole whose excitations are connected by a particle-hole symmetry with a single energy spectrum. 
For each level of the projection, we express generic Hamiltonians which couple to an inversion order parameter ($\phi$).

\subsection{Eight-band model }\label{ss3-1}
\begin{figure}[h]
\vspace{-10pt}
 \centering\includegraphics[scale=0.4]{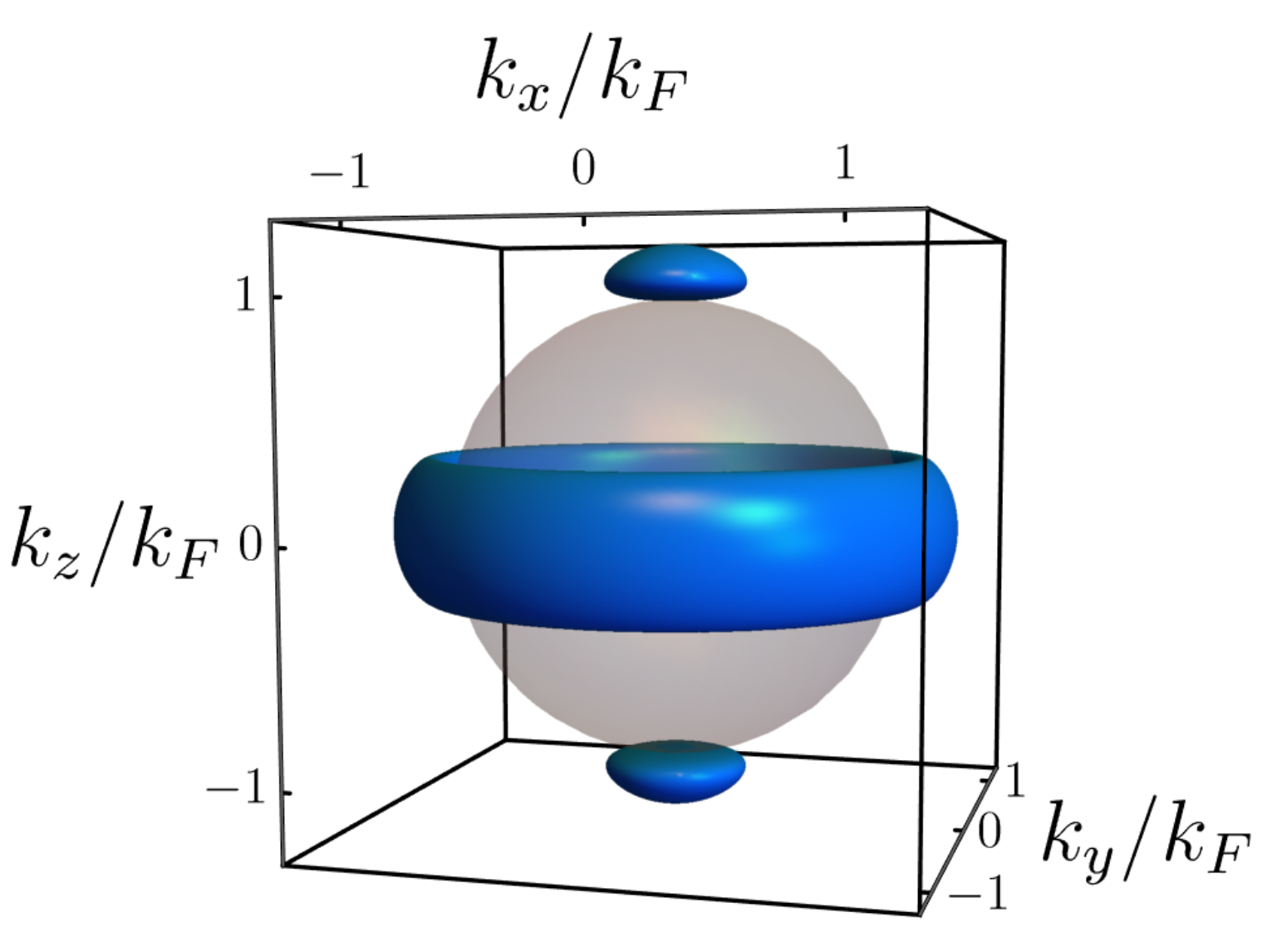}
 \caption{BG-FSs of a chiral TRSB state in a momentum space. The spheroidal and toroidal BG-FSs (blue) are protected by a $Z_{2}$ topological invariant \cite{Agterberg1}. The dimensionless momentum vector ($\vec{k}/k_{F}$) is introduced, where $k_{F}$ is a momentum of the isotropic normal Fermi-surfaces (grey).}\label{sf1}
\end{figure}

Let us start with a $j=3/2$ superconductivity Hamiltonian. Its BdG Hamiltonian is generically written as,  
\begin{eqnarray}
H^{0}&\equiv &\sum_{\vec{k}}\chi_{\vec{k}}^{\dagger}\begin{pmatrix} 
H_{0}(\vec{k}) & \Delta(\vec{k}) \\
\Delta^{\dagger}(\vec{k})  & -H_{0}^{T}(-\vec{k}) 
\end{pmatrix}\chi_{\vec{k}}, \ \ \ \ \chi^{T}_{\vec{k}}=(\xi^{T}_{\vec{k}},\xi_{-\vec{k}}^{\dagger}),
\label{se16}
\end{eqnarray}
where a four-component spinor $\xi_{\vec{k}}=(f_{\vec{k},\frac{3}{2}},f_{\vec{k},\frac{1}{2}},f_{\vec{k},-\frac{1}{2}},f_{\vec{k},-\frac{3}{2}})^{T}$ with a fermionic annihilation operator $f_{\vec{k},\alpha}$. 
The normal and pairing parts, ($H_{0}(\vec{k})$, $\Delta(\vec{k})$) are described by  $4\times4$ matrices. To be specific, we consider a standard Hamiltonian in the literature \cite{Agterberg1}, 
\begin{eqnarray}
 H_{0}(\vec{k}) & = &(\tilde{c}_{0}k^{2}-\epsilon_{\mathrm{F}})\gamma^{0}+\tilde{c}_{1}\sum_{a=1}^{5}d_{a}(\vec{k})\gamma^{a}, \ \ \ \  \Delta(\vec{k})=\Delta_0 (\gamma^{3}+i \gamma^2)i \gamma^{12},\label{se19}
\end{eqnarray}
with $\tilde{c}_{1}=\tilde{c}_2$. 
The overall pairing amplitude $\Delta_{0}$ is a real number, and a chiral time-reversal symmetry breaking (TRSB) pairing $(\gamma^{3}+i \gamma^{2})$ is chosen. 
Zero-energy surfaces in the Brilluion zone are illustrated in Fig. \ref{sf1}.

Now, we introduce a generic Hamiltonian of a BG-FS which couples to an inversion order parameter ($\phi$), 
\begin{eqnarray}
  \delta H^{0} _{\mathrm{Inv}}&=& \phi\  \sum_{R,\vec{k}}\ \chi_{\vec{k}}^{\dagger}\left(\begin{array}{c c}
   {\eta}_{R}(\vec{k})& {\delta}_{R}(\vec{k})\\
   {\delta}^{\ \dagger}_{R}(\vec{k})&- {\eta}^{\ T}_{R}(-\vec{k})
  \end{array}\right)\chi_{\vec{k}}.\label{se30} 
\end{eqnarray}  
 The subscript $R$ is for an irreducible representation whose dimension is specified by $d_{R}$. The $4\times 4$ matrices ($\eta_{R},\delta_{R}$) are  $\eta_{R}(\vec{k})=\sum_{m} a_{m}^{R}\eta_{R,m} (\vec{k})$, $\delta_{R}(\vec{k})=\sum_{m} b^{R}_{m}\delta_{R,m} (\vec{k})$ with coefficients $(a^{R}_{m},b^{R}_{m})$ for $m=1, \cdots d_{R}$. 

 In Table \ref{st2}, odd-parity pairing channels of $O_{h}$ group, $\delta_{R,m}(\vec{k})$, are listed. $O_{h}$ is a higher symmetry group than $C_{i}$, which is a point group of $H^{0}$, hence all odd-parity representations of $O_h$ may mix together and become $A_{u}$ representation of $C_i$.

\subsection{Four-band model }\label{ss3-2}
The construction of a four-band model is standard \cite{Agterberg1,Venderbos}.  
One may construct a four-band model by projecting a eight-band model onto either electron or hole bands of the normal state ($\nu=\pm 1$).  
The condition $\tilde{c}_{0}<\tilde{c}_{1}$ gives well-defined electron and hole bands, so the sign of chemical potential determines a projected Hilbert space.  

To be specific, let us consider electron bands ($\nu=+ 1$) with a positive chemical potential. Its BdG Hamiltonian is,
\begin{eqnarray}
H^{(4)}&\equiv &\sum_{\vec{k}}\tilde{\chi}_{\vec{k}}^{\dagger}\begin{pmatrix} 
H_{+}(\vec{k}) & \Delta_{+}(\vec{k}) \\
\Delta^{\dagger}_{+}(\vec{k})  & -H_{+}^{T}(-\vec{k}) 
\end{pmatrix}\tilde{\chi}_{\vec{k}},\ \ \  \tilde{\chi}_{\vec{k}}^{T}=(\psi_{\vec{k}}^{T},\psi^{\dagger}_{-\vec{k}}).
\end{eqnarray}
A two-component spinor $\psi_{\vec{k}}$ is introduced, and the choice of the spinors is known to have an ambiguity, due to the degeneracy of a normal state energy \cite{Agterberg1}. The superscript $(4)$ denotes a four-band model.

Consider an operator $J_{\vec{k}}\equiv \hat{k}\cdot \vec{J}$ to label two degenerate electron bands as $J=\pm 1/2$. 
By applying a local transformation, spinor basis labeled by $J_{\vec{k}}$ may be transformed to spinors with a pseudo-spin basis, ($\sigma=\pm 1$),
 \begin{eqnarray}
\left(\begin{array}{c}
f_{\vec{k},\sigma=+}\\
f_{\vec{k},\sigma=-}
\end{array}\right)
&\equiv& \exp\left(-i\frac{\sigma^{3}}{2}\phi_{\vec{k}}\right)\exp\left(-i\frac{\sigma^{2}}{2}\theta_{\vec{k}}\right)\left(\begin{array}{c}
f_{\vec{k},J_{\vec{k}}=+\frac{1}{2}}\\
f_{\vec{k},J_{\vec{k}}=-\frac{1}{2}} \nonumber 
\end{array}\right)
\end{eqnarray}
with solid angles in a momentum space ($\theta_{\vec{k}},\phi_{\vec{k}}$). 

The explicit forms of ($H_{+}(\vec{k})$, $\Delta_{+}(\vec{k})$) with a pseudo-spin basis, $\psi_{\vec{k},+}^{T}=(f_{\vec{k},\sigma=+},f_{\vec{k},\sigma=-})$, are 
\begin{eqnarray}
H_{+}(\vec{k})&=& h_0(\vec{k}) \sigma^{0} + \vec{h}(\vec{k})\cdot\vec{ \sigma}, \ \ \ \Delta_{+}(\vec{k})= \psi_{s}(\vec{k})i \sigma^{2},
\end{eqnarray}
with $\sigma^{0}\equiv I_{2\times 2}$, $ \psi_{s}(\vec{k})=\Delta_{0}(\hat{d}_{3}(\vec{k})+i\hat{d}_{2}(\vec{k}))$,  and
\begin{eqnarray}
h_0(\vec{k})&=& E_{0,+}(\vec{k})-\frac{\Delta_{0}^{2}}{2|d(\vec{k})|} (2-\hat{d}_{2}(\vec{k})^{2}-\hat{d}_{3}(\vec{k})^{2}),\ \ \  
h_{\mu}(\vec{k})=\frac{\Delta_{0}^{2}}{|d(\vec{k})|}\left(\frac{\hat{d}_3 (\vec{k})}{\sqrt{3}},\frac{\hat{d}_2 (\vec{k})}{\sqrt{3}},\frac{1-4\hat{d}_5 (\vec{k})}{3}\right). \label{se20}
\end{eqnarray}
where $\hat{d}_{a}(\vec{k})=d_{a}(\vec{k})/|d(\vec{k})|$, and $|d(\vec{k})|=\sqrt{\sum_{a} d_a^2(\vec{k})}$ are used. For the expressions of $(h_0(\vec{k}),h_{\mu}(\vec{k}))$, we keep the corrections up to the second-order in terms of a pairing amplitude. The four-band eigenenergies are 
\begin{eqnarray}
E^{(4)}_{\alpha}(\vec{k})&=&\sqrt{h_{0}(\vec{k})^{2}+|\psi_{s}(\vec{k})|^{2}}+\alpha|\vec{h}(\vec{k})| ,\ \ \ \mathrm{for\ }\  \alpha=\pm1. \label{se27}
\end{eqnarray}
The zero-energy surface state is realized only for $\alpha=-1$, which may dominantly contribute to a low energy physics. 

The unitary transformation from a pseudo-spin basis into an energy eigenvector is well defined,
 \begin{eqnarray}
f_{\vec{k} ,E_{\alpha} ^{(4)}}&=&\left(\frac{\alpha}{2} \right)^{1/2}\begin{pmatrix}
   \sqrt{1+\frac{{0}}{\sqrt{h_{0}^{2}+|\psi_{s}|^{2}}}}\sqrt{\frac{1+\alpha\hat{h}_{3}}{2}}\exp\left(i\frac{(\theta_{s}-\phi_{\bm{h}})}{2}\right) \\\
       \alpha \sqrt{1+\frac{h_{0}}{\sqrt{h_{0}^{2}+|\psi_{s}|^{2}}}}\sqrt{\frac{1-\alpha\hat{h}_{3}}{2}}\exp\left(i\frac{(\theta_{s}+\phi_{\bm{h}})}{2}\right)\\
     -\alpha    \sqrt{1-\frac{h_{0}}{\sqrt{h_{0}^{2}+|\psi_{s}|^{2}}}}\sqrt{\frac{1-\alpha\hat{h}_{3}}{2}}  \exp\left(i\frac{(-\theta_{s}+\phi_{\bm{h}})}{2}\right)\\
            \sqrt{1-\frac{h_{0}}{\sqrt{h_{0}^{2}+|\psi_{s}|^{2}}}}  \sqrt{\frac{1+\alpha\hat{h}_{3}}{2}} \ \exp\left(i\frac{(-\theta_{s}-\phi_{\bm{h}})}{2}\right)
   \end{pmatrix}^{T}\begin{pmatrix}
 f_{\vec{k},\sigma=+}\\
  f_{\vec{k},\sigma=-}\\
   f_{-\vec{k},\sigma=+}^\dagger\\
    f_{-\vec{k},\sigma=-}^\dagger \\
   \end{pmatrix},\nonumber 
   \end{eqnarray}
   with $f_{\vec{k},E_{\alpha}^{(4)}}= f_{-\vec{k},E_{-\alpha}^{(4)}}^\dagger$. We introduce a U(1) phase, $\psi_{s}(\vec{k})\equiv |\psi_{s}(\vec{k})|e^{i\theta_{s}(\vec{k})}$, a normalized pseudo-magnetic field  $\hat{h}(\vec{k})\equiv\vec{h}(\vec{k})/|h(\vec{k})|$, and solid angles of pseudo-magnetic field in a momentum space ($\theta_{h},\phi_{h}$).  \\

We introduce a Hamiltonian which couples to an inversion order parameter ($\phi$) in the four-band model, 
  \begin{eqnarray}
  \delta H  ^{(4)} _{\mathrm{Inv}}&=& \phi \ \sum_{R,\vec{k}}\tilde{\chi}_{\vec{k}}^{\dagger}\left(\begin{array}{c c}
\tilde{ \eta}_{R}(\vec{k})&\tilde{\delta}_{R}(\vec{k})\\
 \tilde{ \delta}^{\dagger}_{R}(\vec{k})&-\tilde{\eta}^{T}_{R}(-\vec{k})
  \end{array}\right)\tilde{\chi}_{\vec{k}}, \label{se31}
  \end{eqnarray}   
The $2\times 2$ matrices $(\tilde{\eta}_{R}$, $\tilde{\delta}_{R})$ are introduced as $\tilde{\eta}_{R}(\vec{k})=\sum_{m} a_{m}^{R}\tilde{\eta}_{R,m} (\vec{k})$, $\tilde{\delta}_{R}(\vec{k})=\sum_{m} b_{m}^{R}\tilde{\delta}_{R,m} (\vec{k})$. Each channel may be explicitly expressed in terms of $4\times4$ matrices $(\eta_{R}$,$\delta_{R})$ in a eight-band model, 
\begin{eqnarray}
\tilde{\eta}_{R}(\vec{k})&=& U_{\vec{k}}^{\dagger}\eta_{R}(\vec{k})U_{\vec{k}}=\sum_{\mu=0,1,2,3}\tilde{\eta}_{R}^{\mu}(\vec{k})\sigma^{\mu},\ \ 
\tilde{\delta}_{R}(\vec{k})= U_{\vec{k}}^{\dagger}\delta_{R}(\vec{k})U_{\vec{k}}^{*}=\sum_{\mu=1,2,3}\tilde{\psi}_{R}^{\mu}(\vec{k})\sigma^{\mu}(i\sigma^{2}),
\end{eqnarray}
by using a transposed matrix,
\begin{eqnarray}
U_{\vec{k}}^{T}&=&\left(
\begin{array}{cccc}
\! -\frac{ \sqrt{3} e^{i \phi_{\vec{k}}  } }{2}\sin \theta_{\vec{k}} \! \!\!  &  \cos \theta_{\vec{k}} \!\! \!  &  \frac{e^{-i \phi_{\vec{k}}}}{2} \sin \theta_{\vec{k}} \!\! &0 \!\! \\
 \! 0\! \!\! &-\frac{e^{i \phi _{\vec{k}} } }{2} \sin \theta_{\vec{k}} \! \! & \cos \theta_{\vec{k}}  \!\! \!  &\frac{ \sqrt{3}e^{-i \phi_{\vec{k}}  }}{2}  \sin \theta_{\vec{k}} \!\! 
\end{array} \nonumber
\right). 
\end{eqnarray}
In Table \ref{st2}, odd-parity pairing channels of the four-band model $\tilde{\delta}_{R,m}$ and the eight-band model $\delta_{R,m}$ are listed.

\subsection{Two-band model }\label{ss3-3}
Similarly, a two-band model may be constructed by projecting a four-band model onto a Hilbert space associated with zero-energy surface states which are manifested in a energy, $E^{(4)}_{-}(\vec{k})=\sqrt{h_{0}(\vec{k})^{2}+|\psi_{s}(\vec{k})|^{2}}-|\vec{h}(\vec{k})|\equiv E_{0}(\vec{k})$. The effective two-band Hamiltonian is,
\begin{eqnarray}
H^{(2)}&\equiv &\sum_{\vec{k}}\Psi_{\vec{k}}^{\dagger} E_{0}(\vec{k})\tau^{z}\ \Psi_{\vec{k}},\ \ \ \Psi_{\vec{k}}^{T}=(f_{\vec{k},E_{-}^{(4)}},f^{\dagger}_{-\vec{k},E_{-}^{(4)}}).
  \end{eqnarray} 
The superscript $(2)$ denotes a two-band model which will be mainly used in the following sections.\\

The Hamiltonian which couples to an inversion order parameter $(\phi)$ in a two-band model is,
 \begin{equation}
    \delta H ^{(2)}_{\mathrm{Inv}}=\phi \  \sum_{\substack{\mu=0,x,y;\\
                  \vec{k},R}}\Psi_{\vec{k}}^{\dagger}\ \rho_{\mu}^{R}(\vec{k})\tau^{\mu}\ \Psi_{\vec{k}}.\label{se32} 
 \end{equation}  
We introduce a real-valued odd-parity function, $\rho_{\mu}^{R}(\vec{k})=\sum a_{\mu,m}^{R}\rho_{\mu}^{R,m}(\vec{k})$. 
Coefficients $a_{\mu,m}$ are labeled by a channel index $\mu$ and an irreducible representation $R$ for $m=1, \cdots d_{R}$. 
  Note that one channel, $\rho_{z}^{R}(\vec{k})$, is forbidden by a particle-hole symmetry.\\

 We derive $\rho_{\mu}^{R}(\vec{k})$ in terms of inversion-symmetry breaking channels, $(\tilde{\eta}_{R}(\vec{k})$, $\tilde{\delta}_{R}(\vec{k}))$, and BdG Hamiltonian parameters, $(h_{0}(\vec{k}),h_{\mu}(\vec{k}),\psi_{s}(\vec{k}))$ in the four-band model,
  \begin{eqnarray}
\rho_{0}^{R}(\vec{k})&=&\left(\tilde{\eta}_{R}^{0}-\frac{h_{0}}{\sqrt{|\psi_{s}|^{2}+h_{0}^{2}}}(\hat{h}\cdot\vec{\tilde{\eta}}_{R})\right)-\mathrm{Re}\left(\frac{ \psi_{s}^*}{\sqrt{|\psi_{s}|^{2}+h_{0}^{2}}}(\hat{h}\cdot\vec{\tilde{\psi}}_{R})\right), \nonumber
\end{eqnarray}\vspace{-10pt}
\begin{eqnarray}
\rho_{x}^{R}(\vec{k})&=&\frac{|\psi_{s}|}{\sqrt{|\psi_{s}|^{2}+h_{0}^{2}}}\frac{\hat{h}_{3}(\hat{h}\cdot \vec{\tilde{\eta}}_{R})-\tilde{\eta}_{R}^{3}}{\sqrt{\hat{h}_{1}^{2}+\hat{h}_{2}^{2}}}-\mathrm{Re}\left(\frac{h_{0}}{\sqrt{|\psi_{s}|^{2}+h_{0}^{2}}}\frac{ \psi_{s}^{*}}{|\psi_{s}|} \frac{\hat{h}_{3}(\hat{h}\cdot\vec{\tilde{\psi}}_{R})-\tilde{\psi}_{R}^{3}}{\sqrt{\hat{h}_{1}^{2}+\hat{h}_{2}^{2}}}\right) +\mathrm{Im}\left(\frac{\psi_{s}^{*}}{|\psi_{s}|}\frac{\hat{h}_{1}\tilde{\psi}_{R}^{2}-\hat{h}_{2}\tilde{\psi}_{R}^{1}}{\sqrt{\hat{h}_{1}^{2}+\hat{h}_{2}^{2}}}\right),\ \ \ \ \ \end{eqnarray}\vspace{-10pt}\begin{eqnarray}
\rho_{y}^{R}(\vec{k})&=& \frac{|\psi_{s}|}{\sqrt{|\psi_{s}|^{2}+h_{0}^{2}}}\frac{\hat{h}_{1}\tilde{\eta}_{R}^{2}-\hat{h}_{2}\tilde{\eta}_{R}^{1}}{\sqrt{\hat{h}_{1}^{2}+\hat{h}_{2}^{2}}}\ -\mathrm{Re}\left(\frac{h_{0}}{\sqrt{|\psi_{s}|^{2}+h_{0}^{2}}}\frac{ \psi_{s}^{*}}{|\psi_{s}|} \frac{\hat{h}_{1}\tilde{\psi}_{R}^{2}-\hat{h}_{2}\tilde{\psi}_{R}^{1}}{\sqrt{\hat{h}_{1}^{2}+\hat{h}_{2}^{2}}}\right) \ -\mathrm{Im}\left(\frac{\psi_{s}^{*}}{|\psi_{s}|}\frac{\hat{h}_{3}(\hat{h}\cdot\vec{\tilde{\psi}}_{R})-\tilde{\psi}_{R}^{3}}{\sqrt{\hat{h}_{1}^{2}+\hat{h}_{2}^{2}}}\right).\ \ \ \ \ \  \nonumber\end{eqnarray}
One can note that both normal, and pairing part in the four-band model mix together in each channel in a two-band model.

\vspace{20pt}
  \begin{table}[h]
\centering
\renewcommand{\arraystretch}{1.35}
\begin{tabular}{C{0.06\linewidth}C{0.43\linewidth}C{0.39\linewidth}}\hline \hline

$R$  &Eight-band model, $\vec{\delta}_{R,m}(\vec{k}) (i\gamma^{12})^\dagger$ &Four-band model, $\vec{\tilde{\delta}}_{R,m}(\vec{k}) (i\sigma^{2})^\dagger$\\ \hline
$A_{1u}$&$\vec{k}\cdot\vec{j},\ \vec{k}\cdot\vec{\mathcal{J}}$&$\vec{k}_{T_{1u}}\cdot \vec{\sigma}$
 \\
 $A_{2u}$&$\vec{k}\cdot\vec{T}$&$\vec{k}_{T_{2u}}\cdot\vec{\sigma}$\\
$E_{u}$&$(D^{4}_{\vec{k},\vec{j}},D^{5}_{\vec{k},\vec{j}}),(D^{4}_{\vec{k},\vec{\mathcal{J}}},D^{5}_{\vec{k},\vec{\mathcal{J}}}),(D^{4}_{\vec{k},\vec{T}},D^{5}_{\vec{k},\vec{T}})$&$(D^{4}_{\vec{k}_{T_{1u}},\vec{\sigma}},D^{5}_{\vec{k}_{T_{1u}},\vec{\sigma}}),(D^{4}_{\vec{k}_{T_{2u}},\vec{\sigma}},D^{5}_{\vec{k}_{T_{2u}},\vec{\sigma}})$\\
$T_{1u}$&$\vec{k}\times\vec{j},\ \vec{k}\times \vec{\mathcal{J}},(D^{1}_{\vec{k},\vec{T}},D^{2}_{\vec{k},\vec{T}},D^{3}_{\vec{k},\vec{T}})$&$\vec{k}_{T_{1u}}\times\vec{\sigma},(D^{1}_{\vec{k}_{T_{2u}},\vec{\sigma}},D^{2}_{\vec{k}_{T_{2u}},\vec{\sigma}},D^{3}_{\vec{k}_{T_{2u}},\vec{\sigma}})$
 \\
$T_{2u}$&$\vec{k}\times\vec{T},\vec{k}\gamma^{45},(D^{1}_{\vec{k},\vec{j}},D^{2}_{\vec{k},\vec{j}},D^{3}_{\vec{k},\vec{j}}),(D^{1}_{\vec{k},\vec{\mathcal{J}}},D^{2}_{\vec{k},\vec{\mathcal{J}}},D^{3}_{\vec{k},\vec{\mathcal{J}}})$&$\vec{k}_{T_{2u}}\times \vec{\sigma}, k_{A_{1u}}\vec{\sigma}, (D^{1}_{\vec{k}_{T_{1u}},\vec{\sigma}},D^{2}_{\vec{k}_{T_{1u}},\vec{\sigma}},D^{3}_{\vec{k}_{T_{1u},\vec{\sigma}}})$\\
\hline
\hline
\end{tabular}
\caption{ The odd-parity pairing matrices of $O_{h}$ symmetry with \textit{eight-}, and \textit{four-}band model ($\delta_{R,m}(\vec{k}) ,\tilde{\delta}_{R,m}(\vec{k}) $).
 We introduce $4\times 4$ matrices $ \vec{j}\equiv \frac{2}{\sqrt{5}}J_{i},\  \vec{\mathcal{J}}\equiv\frac{-41}{6\sqrt{5}}J_{i}+\frac{2\sqrt{5}}{3}J_{i}^3, \ \vec{T}\equiv \frac{1}{\sqrt{3}}\left(\left\{J_{1},J_{2}^{2}-J_{3}^{2}\right\},\cdots \right)$. The basis functions of $O_{h}$ group, $k_{A_{2u}}\equiv k_{x}k_y k_z,\vec{k}_{T_{1u}}\equiv  ak_{i}+bk_i^3,\vec{k}_{T_{2u}}\equiv (k_x(k_y^2-k_z^2),\cdots)$ are used with constants $a,b$. Here $\cdots$ denotes a cyclic permutation. Bilinear operations of vectors $D^{a}_{\vec{u},\vec{v}}\equiv \Lambda_{a}^{ij}u_{i}v_{j}$ are defined by five $3\times 3$ symmetric Gell-Mann matrices, $\Lambda_{a}\equiv \frac{\sqrt{3}}{2}(\lambda_{6},\lambda_{4},\lambda_{1},\lambda_3,-\lambda_8)$. \cite{Georgi}
 }
\label{st2}
\end{table}

\newpage
\section{mean-field analysis }\label{s5}
Let us consider a BdG Hamiltonian of a two-band model (See section \ref{ss3-3}),
\begin{eqnarray}
H _{0}^{\mathrm{eff}}&= &\sum_{\vec{k}} \Psi^{\dagger}_{\vec{k}}\  E_0(\vec{k}) \tau^z \ \Psi_{\vec{k}}, \label{se36}\end{eqnarray}\vspace{-12pt}
\begin{eqnarray}
H_{\mathrm{int}}^{\mathrm{eff}}&=& -\frac{1}{2} \sum_{\mu,\nu;\vec{k},\vec{k}'} g_{\mu\nu}V_{\mu \nu}(\vec{k},\vec{k}') (\Psi^{\dagger}_{\vec{k}} \tau^{\mu} \Psi_{\vec{k}}) ( \Psi^{\dagger}_{\vec{k}'} \tau^{\nu} \Psi_{\vec{k}'} ). \label{se37}
\end{eqnarray}
We focus on a separable potential $V_{\mu\nu}(\vec{k},\vec{k}') = \rho_{\mu}(\vec{k}) \rho_{\nu}(\vec{k}')$ with $\rho_{\mu}(\vec{k})=-\rho_{\mu}(-\vec{k})$ at a BG-FS. For simplicity, we consider only two channels ($\tau^0,\tau^x$) and three coupling constants ($g_{00}, g_{x0}, g_{xx}$) with $g_{x0}=g_{0x}$. Its generalization including other channels ($\tau^y,\tau^z$) is straightforward. The superscript $\textit{eff}$ would dropped hereafter.

The mean-field approximation with the ansatz
\begin{eqnarray}
 \langle   \Psi^{\dagger}_{\vec{k}}  \tau^{\mu}\Psi_{\vec{k}}\rangle _{\mathrm{MF}} \equiv \phi \,c^{\mu}_{\vec{k}}, 
\end{eqnarray}
gives the mean field Hamiltonian, 
\begin{eqnarray}
H_{\mathrm{MF}}&=&H_{0}-\phi\sum_{\mu, \vec{k}}d_{\mu}\rho_{\mu}(\vec{k}) \Psi^{\dagger}_{\vec{k}}  \tau^{\mu}\Psi_{\vec{k}} +\frac{\phi^{2}}{2}\sum_{\mu,\nu}d_{\mu} g_{\mu\nu}^{-1}d_{\nu}. 
\end{eqnarray}
With an inversion-symmetry breaking order parameter $\phi$, we have
\begin{eqnarray}
d_{\mu} \equiv \sum_{\nu ,\vec{k}}g_{\mu\nu} \rho_{\nu}(\vec{k}) c^{\nu}_{\vec{k}}, \ \ \ \mathrm{for}\  \mu,\nu=0,x .\nonumber 
\end{eqnarray}
Generically, $g_{\mu\nu}$ is non-singular with a well-defined inverse matrix $g_{\mu\nu}^{-1}$.
The mean-field Hamiltonian may be diagonalized by a unitary transformation, 
\begin{eqnarray}
H_{\mathrm{MF}} &=& \sum_{\vec{k}}  \tilde{\Psi}_{\vec{k}}^{\dagger} \Big( -\phi d_{0}\rho_{0}(\vec{k}) \tau^0+\sqrt{ E_0(\vec{k})^2 +\phi^2 d_{x}^2\rho_{x}(\vec{k})^2 }\tau^z\Big) \tilde{\Psi}_{\vec{k}} +\frac{\phi^{2}}{2}\sum_{\mu,\nu}d_{\mu} g_{\mu\nu}^{-1}d_{\nu},  \nonumber
\end{eqnarray}
where $\tilde{\Psi}_{\vec{k}}^{T}=(\gamma_{+}(\vec{k}),\gamma_{-}^{\dagger}(-\vec{k}))$. 
Using an anticommutation relation, $\{\gamma_{\alpha}(\vec{k}),\gamma_{\beta}^{\dagger}(\vec{k}')\}=\delta_{\alpha,\beta} \delta_{\vec{k},\vec{k}'}$, one can simplify it, 
\begin{eqnarray}
H_{\mathrm{MF}} &=&\sum_{\vec{k},\alpha=\pm} E(\vec{k};\phi) \,\gamma^{\dagger}_{\alpha}(\vec{k})  \gamma_{\alpha}  (\vec{k}) +\mathcal{E}_0(\phi),
\end{eqnarray}
where
\begin{eqnarray}
E(\vec{k};\phi) &=&  \sqrt{E_0(\vec{k})^2 + \phi^2 d_{x}^{\:2} \rho_x(\vec{k})^2 } -\phi d_0\rho_0(\vec{k}), 
\end{eqnarray}
and
\begin{eqnarray}
\mathcal{E}_0(\phi) &\equiv& - \sum_{\vec{k}} \sqrt{E_0(\vec{k})^2 + \phi^2 d_{x}^{\:2} \rho_x(\vec{k})^2 } + \frac{\phi^{2}}{2}\sum_{\mu,\nu}d_{\mu} g_{\mu\nu}^{-1}d_{\nu}.\nonumber 
\end{eqnarray}
The ground state energy is
\begin{eqnarray}
 E_{G}^{\mathrm{MF}}[\phi]&=&\mathcal{E}_{0}(\phi)+2 \sum_{\vec{k} \in M^-}E(\vec{k};\phi). 
 \end{eqnarray}
Note that the ground state energy depends on both $(\rho_{0}(\vec{k}),\rho_{x}(\vec{k}))$, and the negative energy manifold $M^-$ is specified by the condition $E(\vec{k};\phi)<0$.

The instability of inversion symmetry may be understood by the ground energy difference, 
 \begin{eqnarray}
 \Delta E_{G}^{\mathrm{MF}}[\phi]\equiv  E_{G}^{\mathrm{MF}}[\phi]- E_{G}^{\mathrm{MF}}[0]&=&-\sum_{\vec{k}}\left(\sqrt{E_0(\vec{k})^2 + \phi^2 d_{x}^{\:2} \rho_x(\vec{k})^2 }-|E_0(\vec{k})|\right)+2 \sum_{\vec{k} \in M^-} E(\vec{k};\phi)+\frac{\phi^{2}}{2}\sum_{\mu,\nu}d_{\mu} g_{\mu\nu}^{-1}d_{\nu}. \nonumber \label{se49}
\end{eqnarray}
The BCS type instability manifests in the first term on the right hand side as shown in the main-text, unless fine-tuned with $d_x \rho_{x}(\vec{k})=0$.  \

A manifold of zero-energy excitation of a centrosymmetric BG-FS after a inversion-symmetry breaking is determined by the condition, $|E(\vec{k};\phi)|=0$, which is equivalent to
\begin{eqnarray}
E_{0}(\vec{k})^{2}+\phi^{2}d_{x}^{2}\rho_{x}(\vec{k})^{2}&=&\phi^{2}d_{0}^{2}\rho_{0}(\vec{k})^{2}.
\end{eqnarray}
The condition constrains the three variables, $\vec{k} = (k_x, k_y, k_z)$, and a surface in the momentum space appears. Thus, a Fermi-surface of BG quasi-particles generically appear. 
On the other hand, if fine-tuned to satisfy $|d_{x}\rho_{0}(\vec{k})|\le |d_{0}\rho_{x}(\vec{k})|$ for all $\vec{k}$, then the two conditions are necessary, 
\begin{eqnarray}
E_{0}(\vec{k})=0, \quad |d_{x}\rho_{0}(\vec{k})|= |d_{0}\rho_{x}(\vec{k})|,
\end{eqnarray}
which cannot give a Fermi surface generically. For example, if we start with $\rho_0(\vec{k})=0$, then a Fermi-surface of BG quasiparticles is absent. 
%


The mean field free energy is 
\begin{eqnarray}
F_{\mathrm{MF}}[\phi]&=& -T \log ({\rm Tr}(e^{-H_{\mathrm{MF}}[\phi]/T}) )=-2T\ \sum_{\vec{k}}\ \log  \left(2\cosh \frac{ E(\vec{k};\phi) }{2T}\right) +\frac{\phi^{2}}{2}\sum_{\mu,\nu}d_{\mu} g_{\mu\nu}^{-1}d_{\nu}.
\end{eqnarray}

%
%
%
%
 
By differentiating $F_{\mathrm{MF}}$ with respect to $\phi d_{\mu}$, one can obtain,
\begin{eqnarray}
\frac{\partial F_{\mathrm{MF}}}{\partial (\phi d_{\mu})}=-A_{\nu}(\phi;T)+\sum_{\nu}g_{\mu \nu}^{-1} \phi d_{\nu}=0,\ \ A_{\mu}(\phi;T)\equiv \sum_{\vec{k}}\tanh\frac{ E(\vec{k};\phi) }{2T}\:\frac{\partial E(\vec{k};\phi) }{\partial (\phi d_{\mu})},
\end{eqnarray}
 which gives simple self-consistent equations,
 \begin{eqnarray}
\phi\: d_{\mu }=\sum_{\nu}g_{\mu\nu}A_{\nu}(\phi;T).\label{se42}
\end{eqnarray} 

\subsection{Determination of phase boundaries}

The inversion symmetry breaking phase boundary ($T^{*}$) for given coupling constants ($g_{00},g_{x0},g_{xx}$) is determined by the self-consistent equations, Eq. (\ref{se42}) with taking a limit of $\phi\rightarrow 0$,
\begin{eqnarray}
1=g_{00}C_{0}(T)+\frac{d_{x}}{d_0}C_{x}(T^{*}), \ 1=\frac{d_{0}}{d_x}C_{0}(T^{*})+g_{xx}C_{x}(T^{*}), \ \ \  C_{\mu}(T)\equiv \lim_{\phi\rightarrow 0}\left(\frac{ A_{\mu}(\phi;T)}{\phi d_{\mu}}\right).\label{se43}
\end{eqnarray}
Note that the phase boundary determination method that we apply in the mean-field analysis is only reliable for a second order transition, otherwise we require the  full mean field calculations. 

 Equating two equations, Eq. (\ref{se43}), and performing a second derivative test may give a compact form of $T^{*}$,
  \begin{eqnarray}
( g_{00}-C_{0}(T^{\:*})^{-1})(g_{xx}-C_{x}(T^{\:*})^{-1})&=&g_{x0}^{\:2},\ \ \ \mathrm{with\ \ \  }g_{\mu \mu}-C_{\mu}(T^{\:*})^{-1}<0 .\label{se46}
 \end{eqnarray}
A symmetry-broken phase $(\phi\neq 0)$ is stabilized for $T<T^{*}(\{g_{\mu\nu}\})$, otherwise a symmetric phase $(\phi=0)$ is stabilized. The temperature dependence of $C_{\mu}(T)$ is,
\begin{eqnarray}
C_{0}(T)=\frac{1}{2T} \sum_{\vec{k}}\rho_{0}(\vec{k})^2 \ \mathrm{sech}^2 \frac{ E_{0}(\vec{k})}{2T}, \quad C_{x}(T)=\sum_{\vec{k}}\rho_{x}(\vec{k})^2 \ \frac{\tanh \frac{ E_{0}(\vec{k})}{2T}}{E_{0}(\vec{k})} . \nonumber 
\end{eqnarray}

\begin{figure}[tb]
\begin{center}
\ \qquad\qquad (a) \ \  \ \ \qquad\qquad \qquad\qquad \qquad \qquad \qquad\ \ \ \quad\ \ (b) \qquad \qquad\qquad \qquad\qquad \qquad\qquad \qquad\qquad \qquad \ \  \ \ \  \\ 
\includegraphics[scale=0.5]{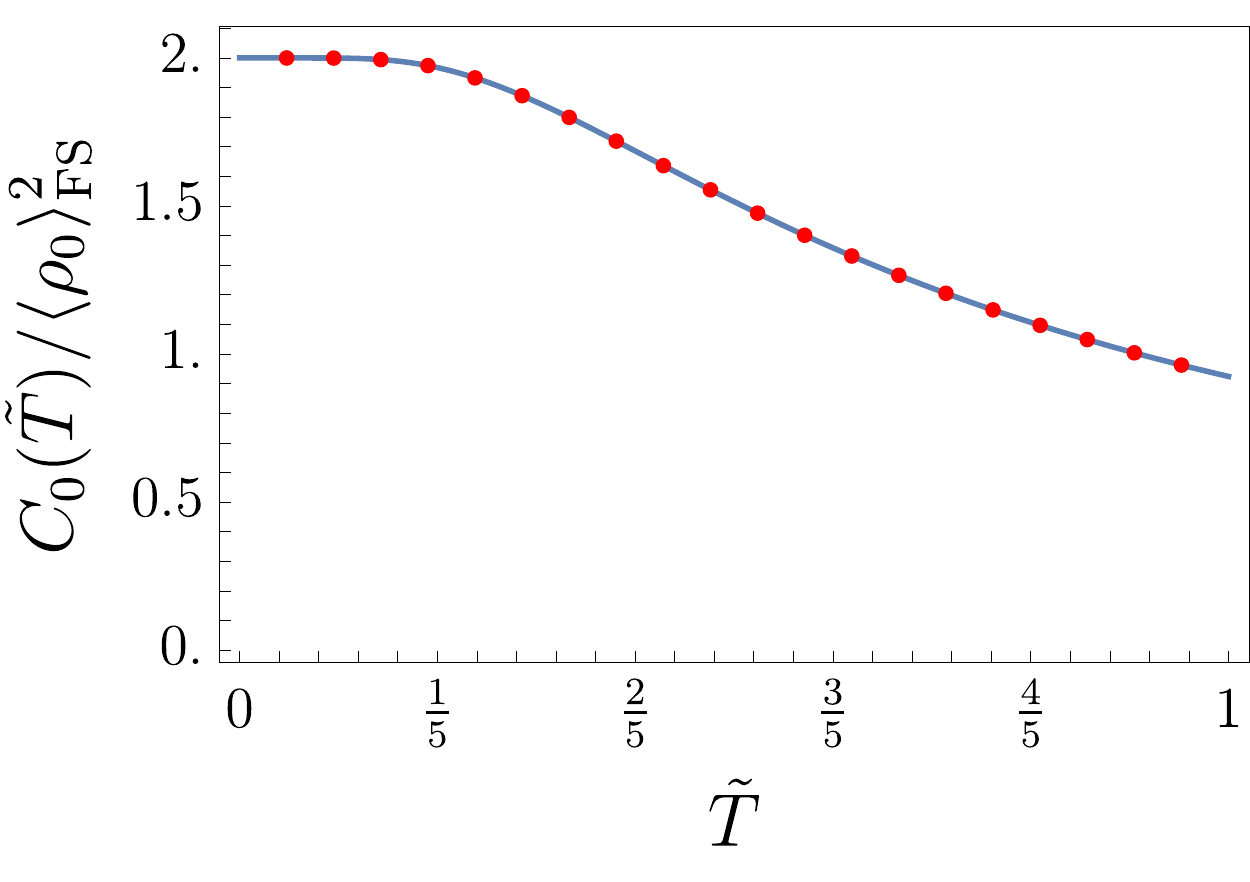}\ \ \ \ \ \ \ \ 
\includegraphics[scale=0.5]{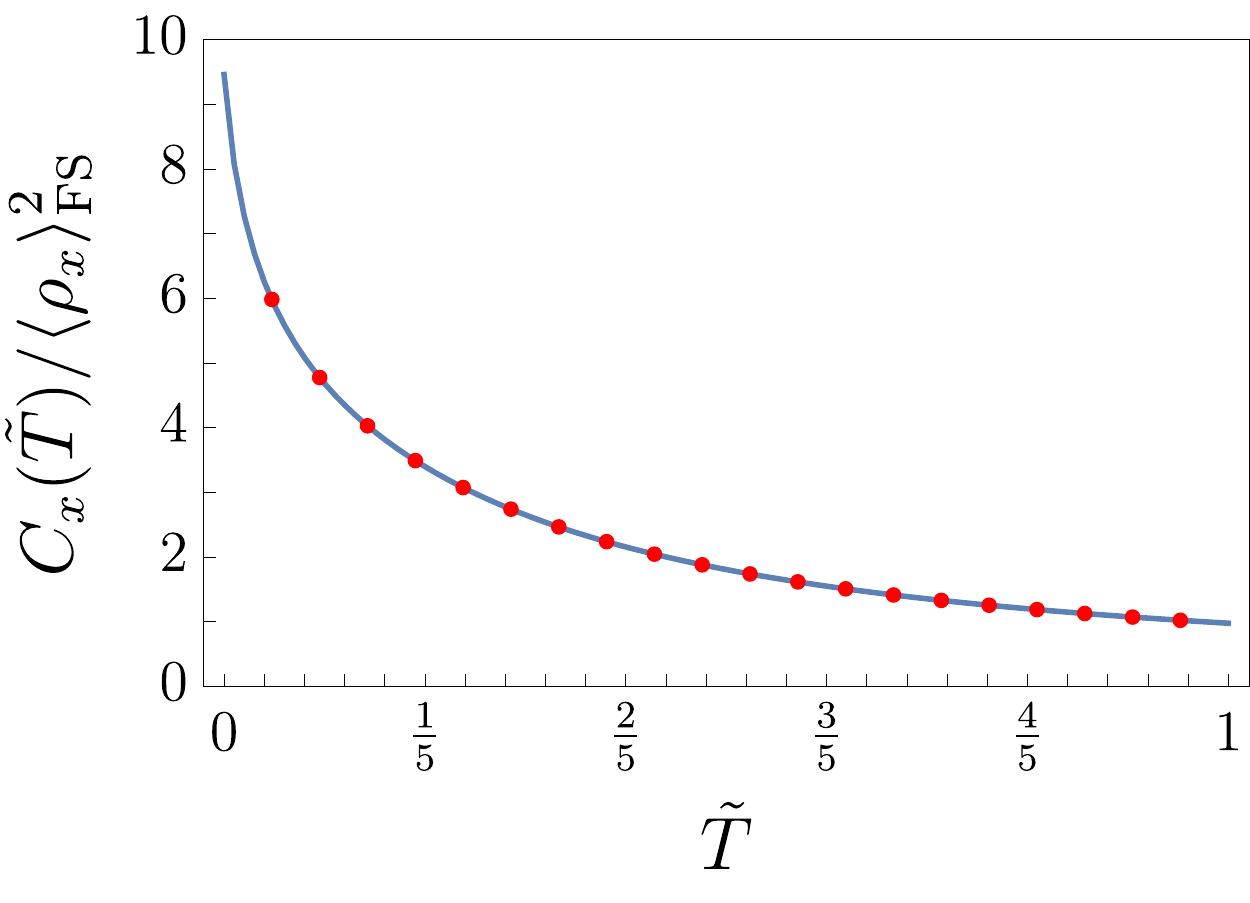}

\caption{Temperature dependence of $C_{0}(\tilde{T})/\langle\rho^{2}_{0}\rangle_{\mathrm{FS}}$, and $C_{x}(\tilde{T})/\langle\rho^{2}_{x}\rangle_{\mathrm{FS}}$ (a,b). The dimensionless energy scale $\tilde{T}=T/\Lambda$ where UV energy cutoff scale $\Lambda$ is used for $x$-axis.  $\lim_{T\rightarrow0} C_{x}(\tilde{T})= \infty$ indicates an inversion symmetry instability. } 
\label{sf2}
\end{center}
\end{figure}

To simplify further, we employ the conventional low temperature approximation,  $\langle N(\tilde{E}) \, \rho_{\mu}( \Omega_{\vec{k}})^{2} \rangle_{\Omega}\simeq \langle N_{\mathrm{F}}(0) \, \rho_{\mu}( \Omega_{\vec{k}})^{2} \rangle_{\Omega}\equiv \langle\rho_{\mu}^{2} \rangle_{\mathrm{FS}}$, and find the equations
\begin{eqnarray}
C_{0}(\tilde{T})\simeq \frac{\left\langle \rho_{0}^2\right\rangle_{\mathrm{FS}}}{2\tilde{T}} \int^{1}_{-1} d\tilde{E} \ \mathrm{sech}^2 \frac{ \tilde{E}}{2\tilde{T}}, \quad C_{x}(\tilde{T})\simeq 
\left\langle \rho_{x}^2\right\rangle_{\mathrm{FS}} \int^{1}_{-1} d\tilde{E} \ \frac{\tanh \frac{\tilde{ E}}{2\tilde{T}}}{\tilde{E}}.\label{se47}
\end{eqnarray}
where dimensionless variables $(\tilde{E},\tilde{T})=(E,T)/\Lambda$ are introduced with a high energy cutoff $\Lambda$, for example a band-width. A typical temperature dependence of $C_{\mu}(\tilde{T})$ is illustrated in Fig. \ref{sf2}. The details of a microscopics ($E_{0}(\vec{k}),\rho_{\mu}(\vec{k})$) are manifested in $\left\langle \rho_{\mu}^2\right\rangle_{\mathrm{FS}}$, but an inversion instability is a generic behaviour, due to the property $\lim_{T\rightarrow0} C_{x}(\tilde{T})= \infty$. At zero temperature, all inversion order parameters $(\phi, d_{0},d_{x})$ condense with non-zero values in the presence of a non-vanishing inter-coupling constant.

\begin{figure}[h]
\begin{center}
(a) \textit{Case 1)} $\rho_{0}(\vec{k})=k_{y}/|k|$,  $\rho_{x}(\vec{k})=k_{x}/|k|$ \\ \vspace{15pt}
\includegraphics[scale=0.28]{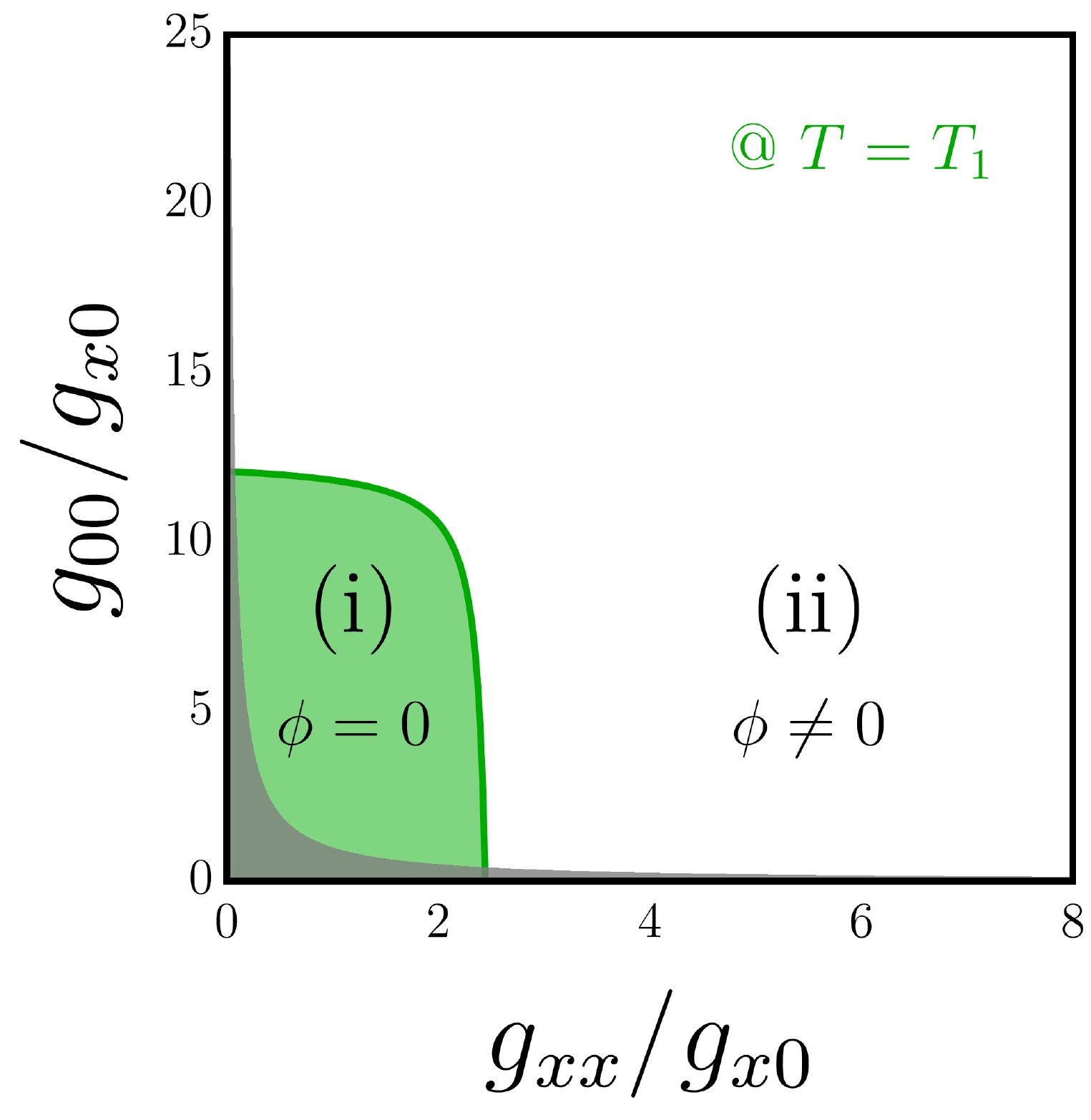}\ \ 
 \includegraphics[scale=0.28]{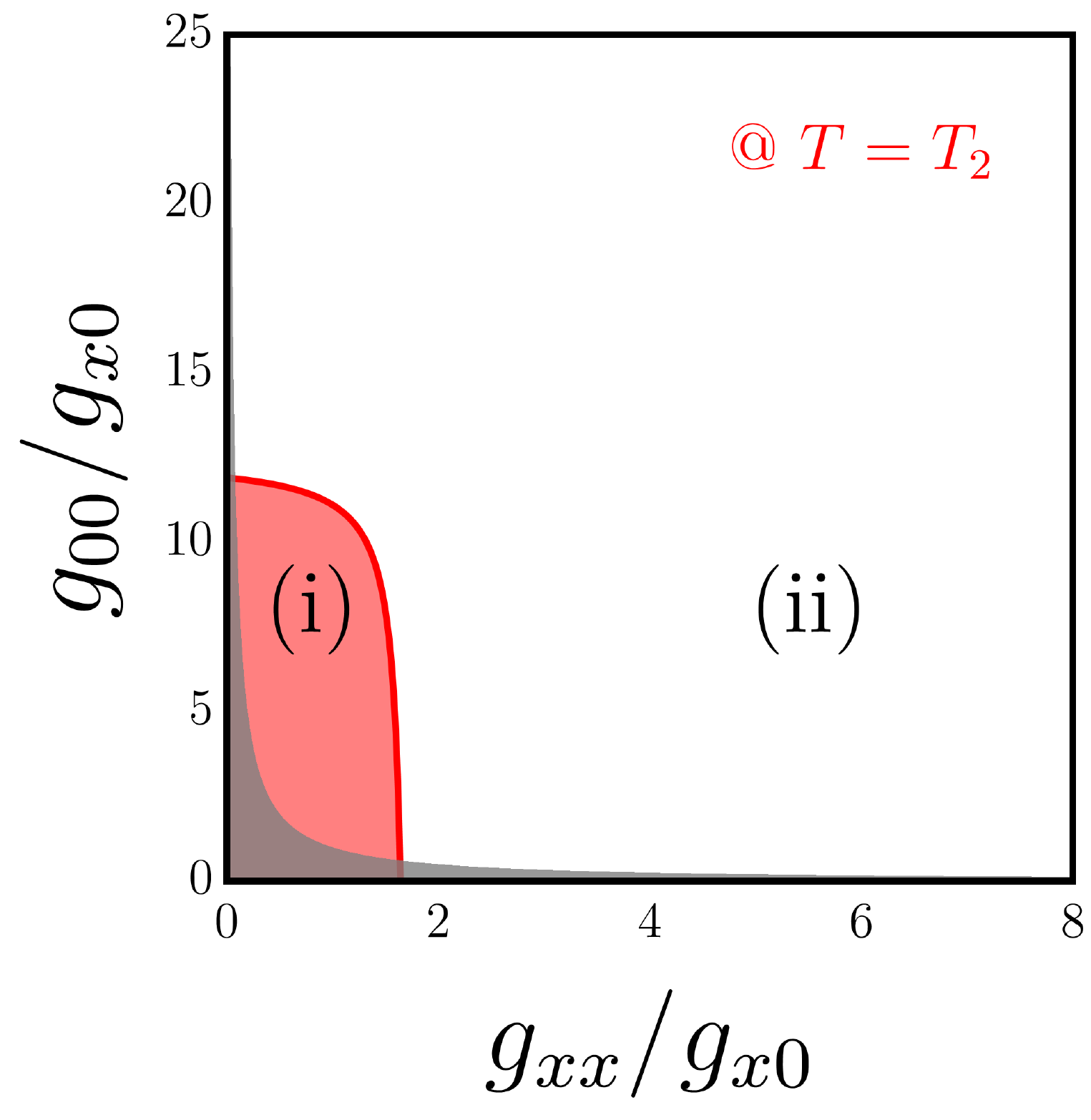}\ \ 
 \includegraphics[scale=0.28]{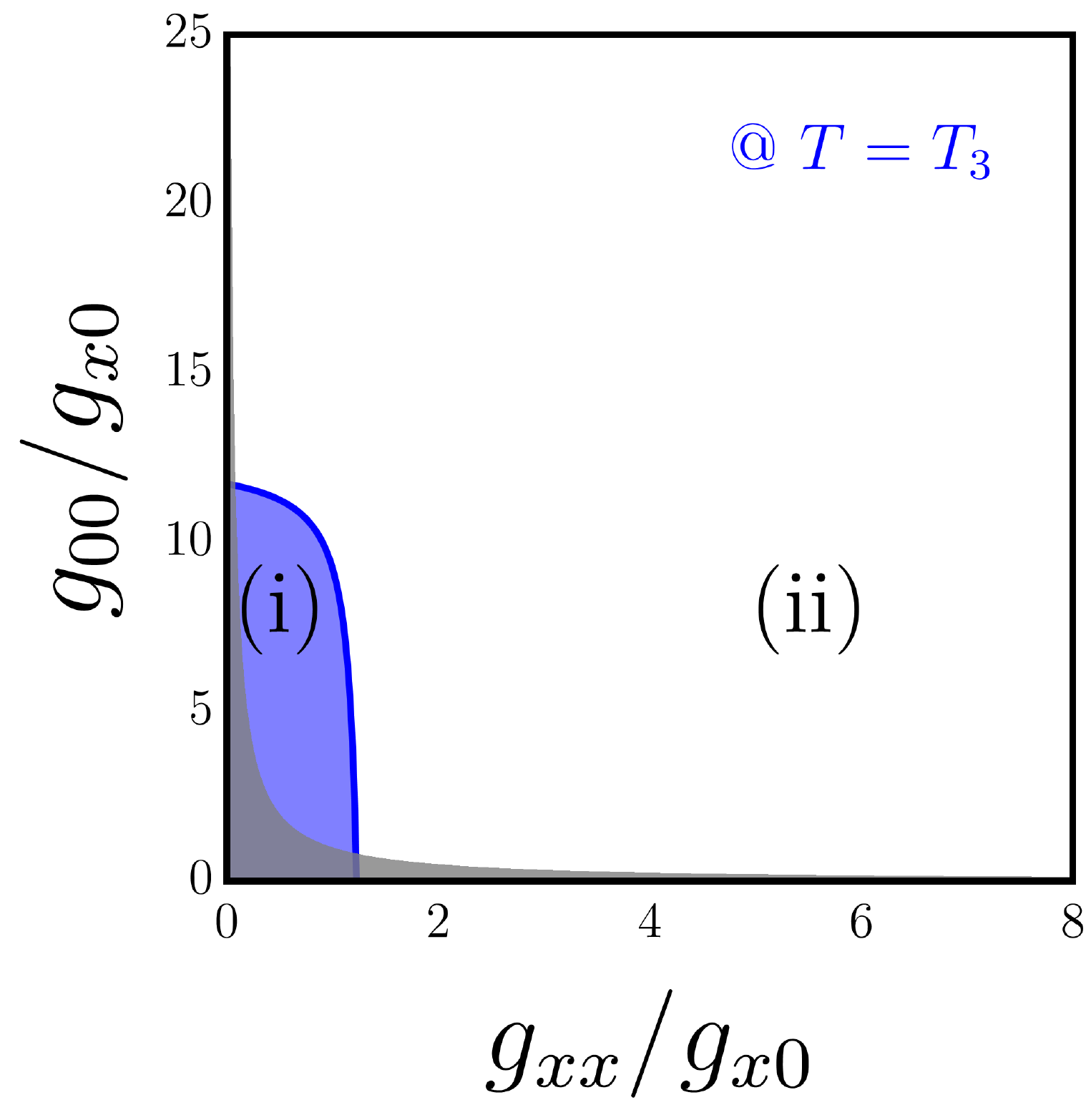}\ \ 
\vspace{15pt}\\

(b) \textit{Case 2)} $\rho_{0}(\vec{k})=\rho_{x}(\vec{k})=k_{x}/|k|$  \\ \vspace{15pt}
\includegraphics[scale=0.28]{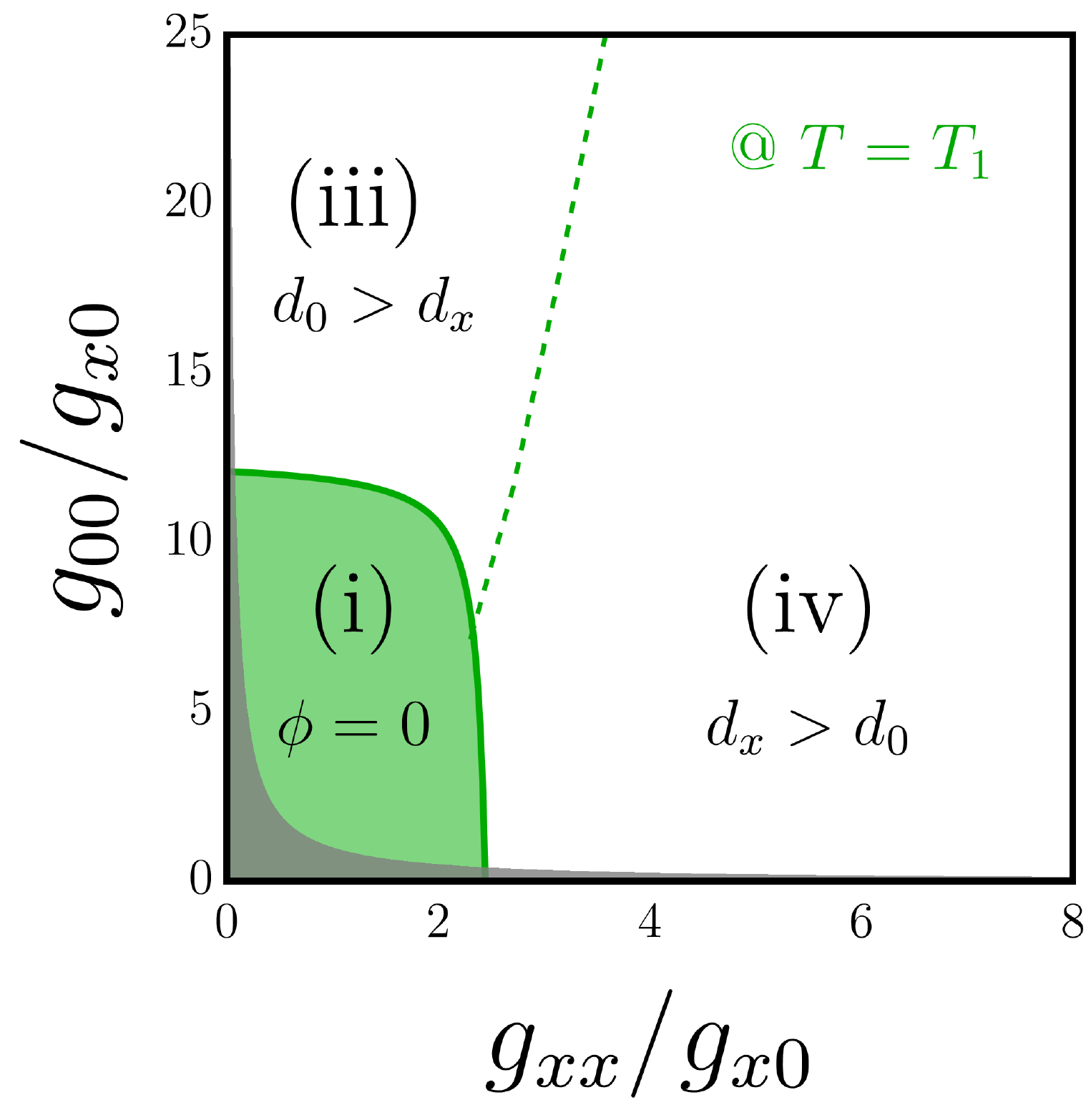}\ \ 
 \includegraphics[scale=0.28]{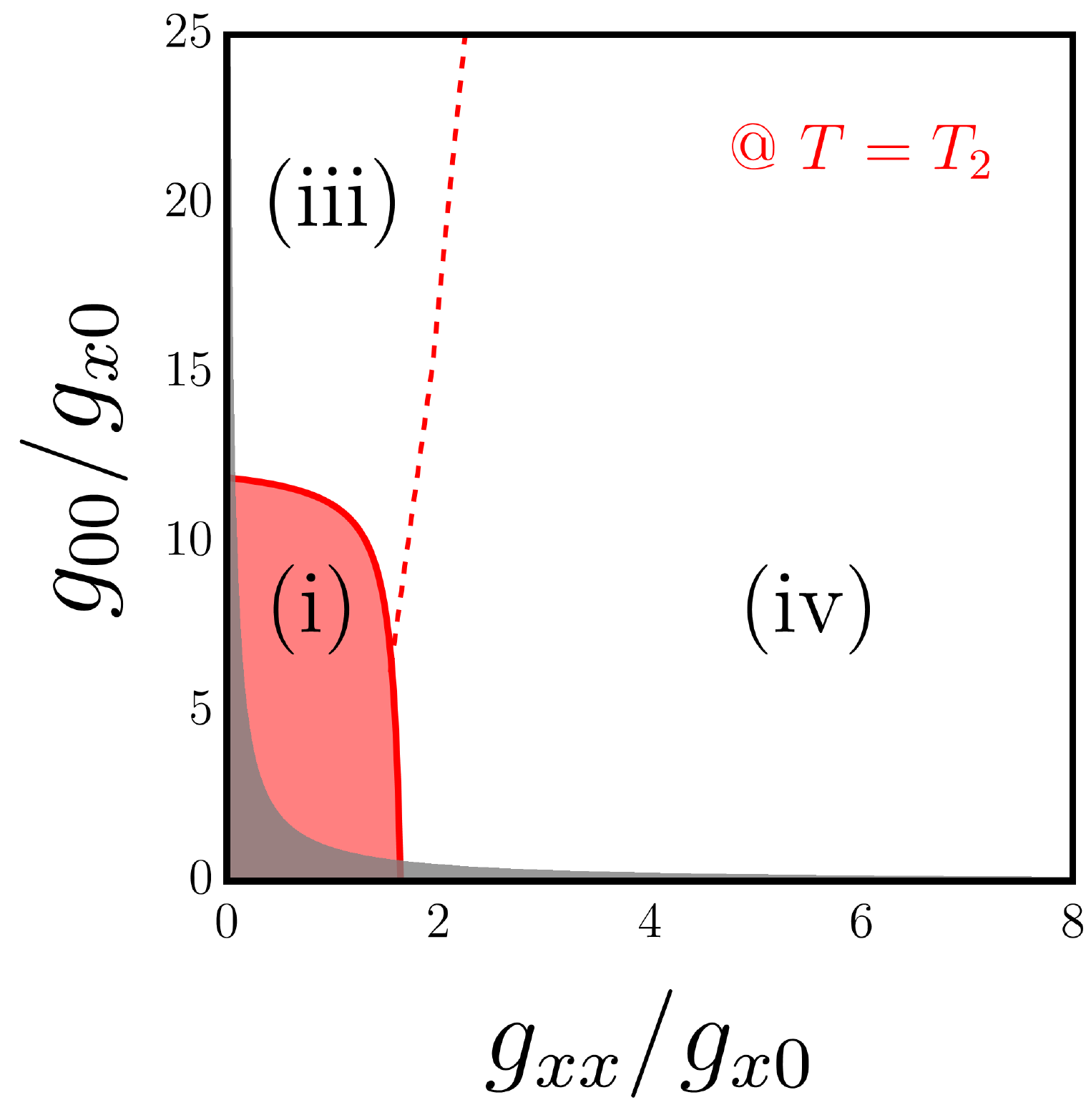}\ \ 
 \includegraphics[scale=0.28]{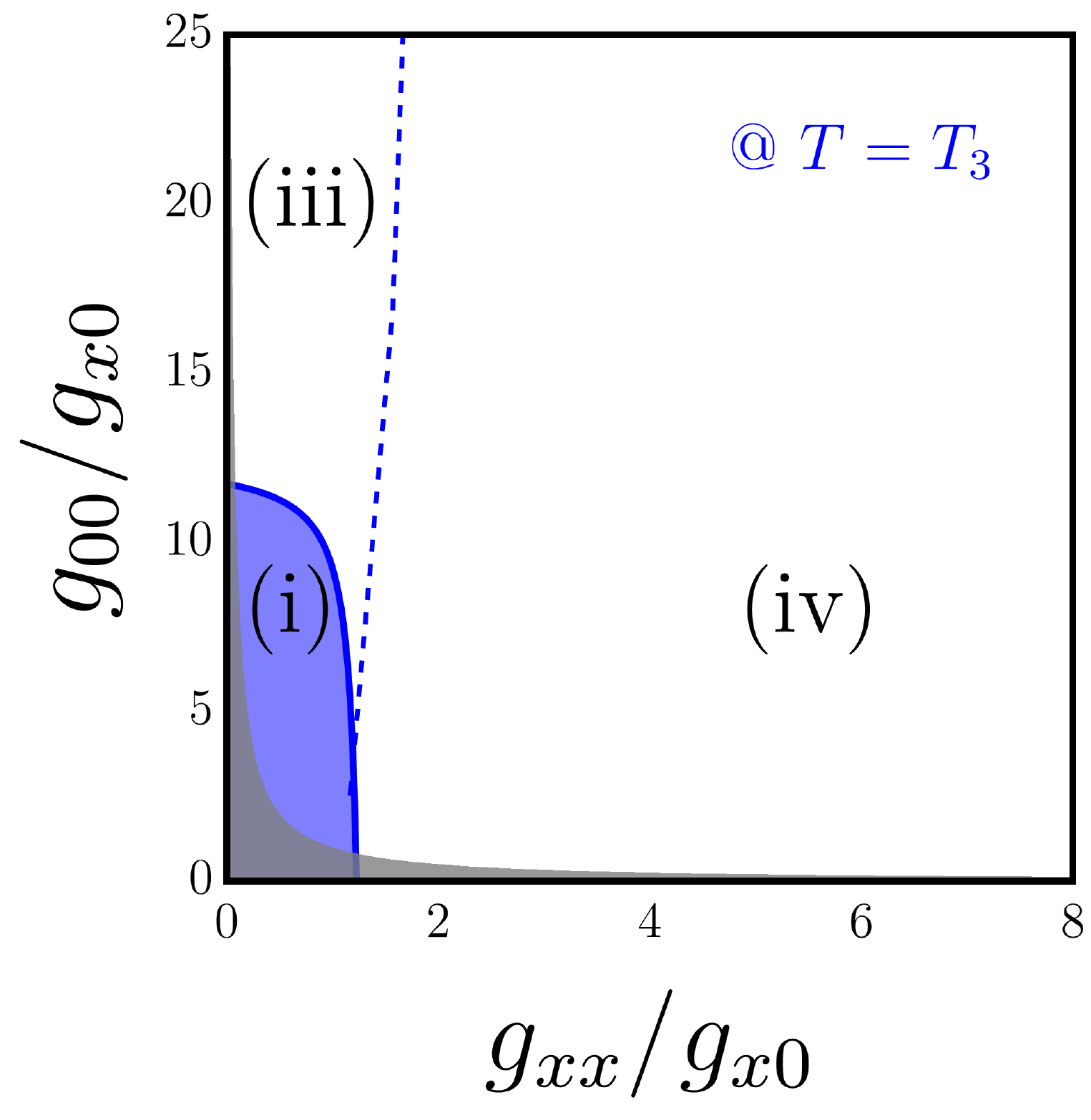}\ \  \\
  \vspace{10pt}
(c) Zero-energy manifolds\\ \vspace{15pt}
 \ \ \ \ \ \ \includegraphics[scale=0.19]{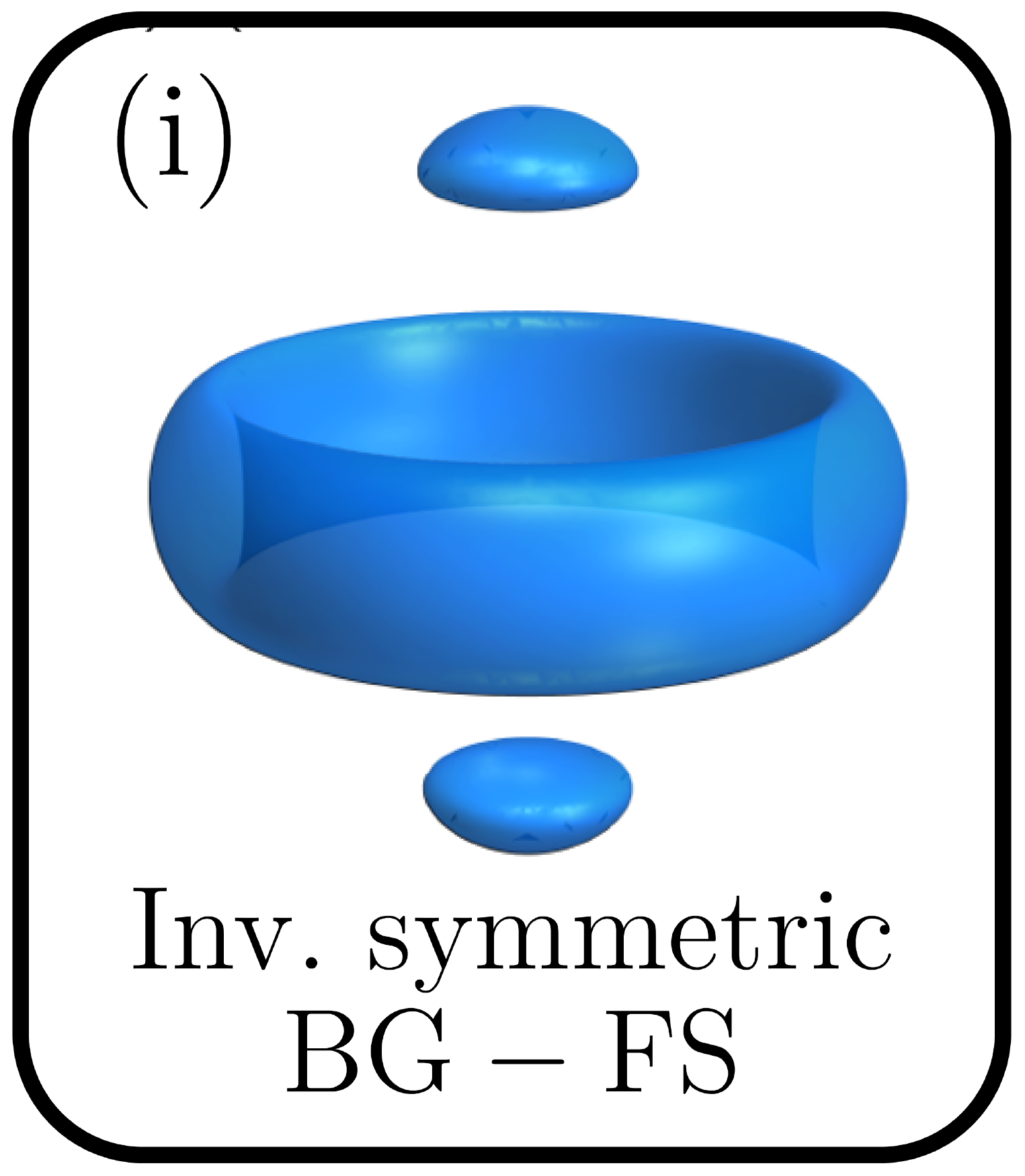} \ \ 
 \includegraphics[scale=0.19]{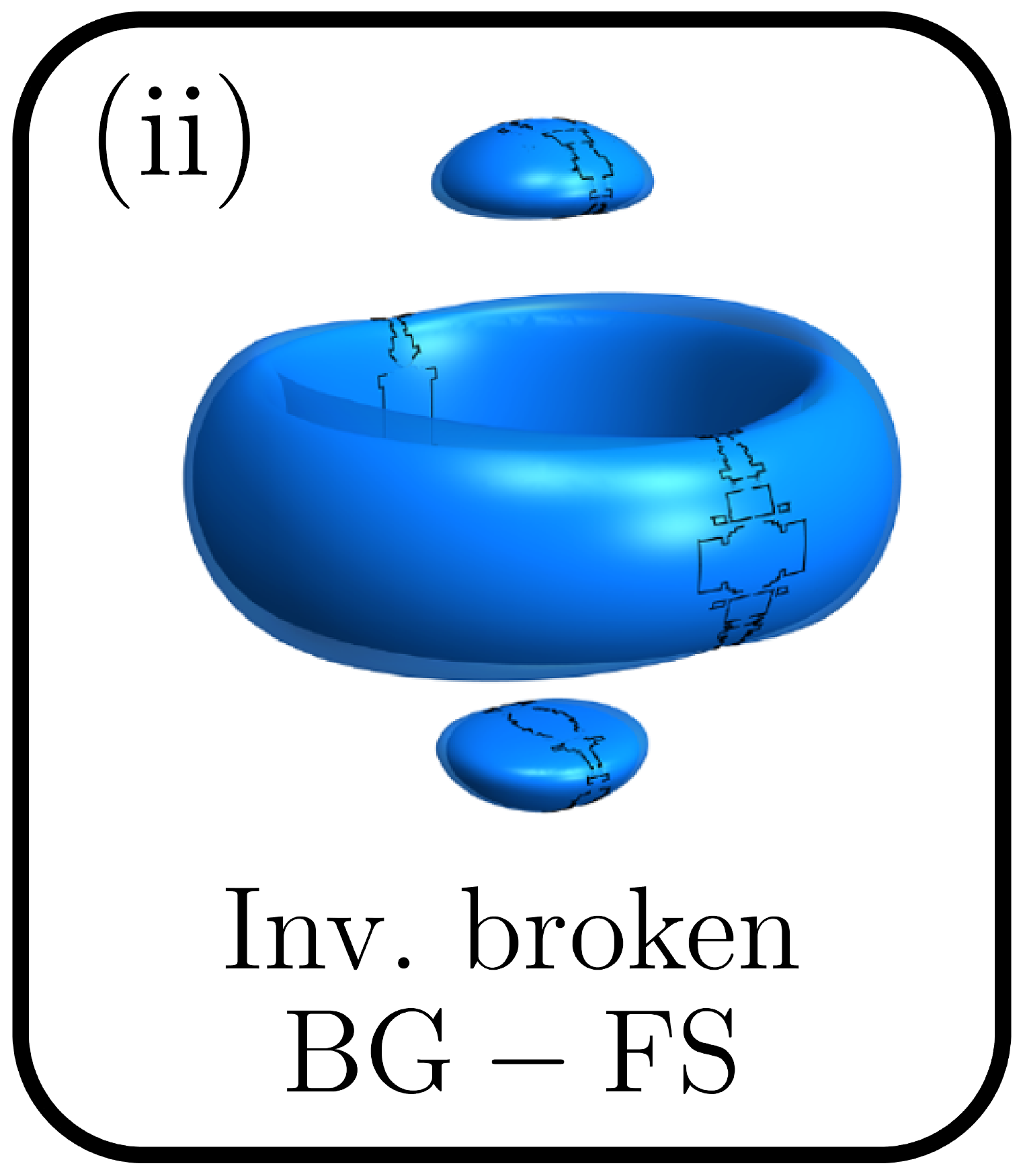} \ \ 
  \includegraphics[scale=0.19]{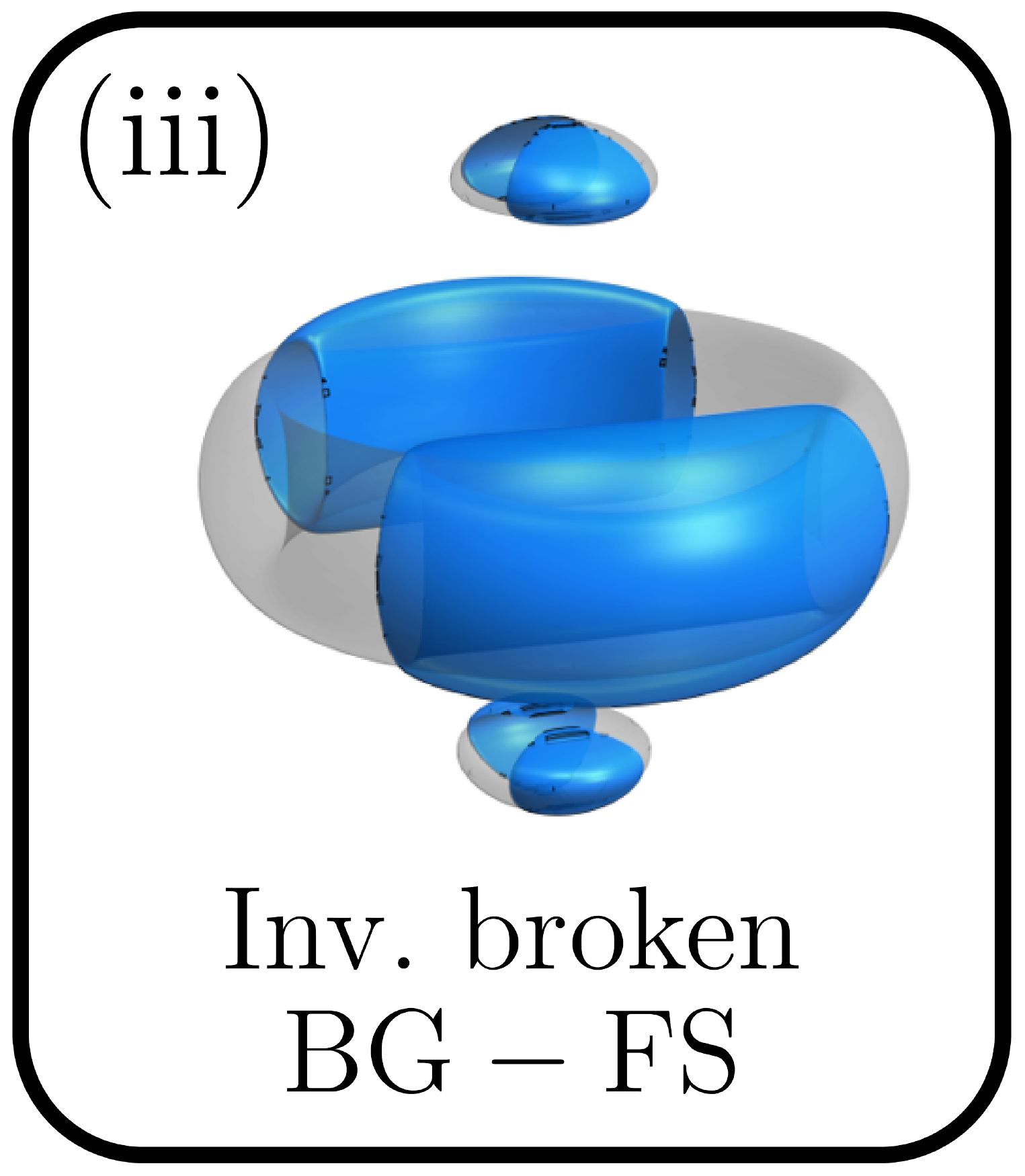} \ \ 
   \includegraphics[scale=0.19]{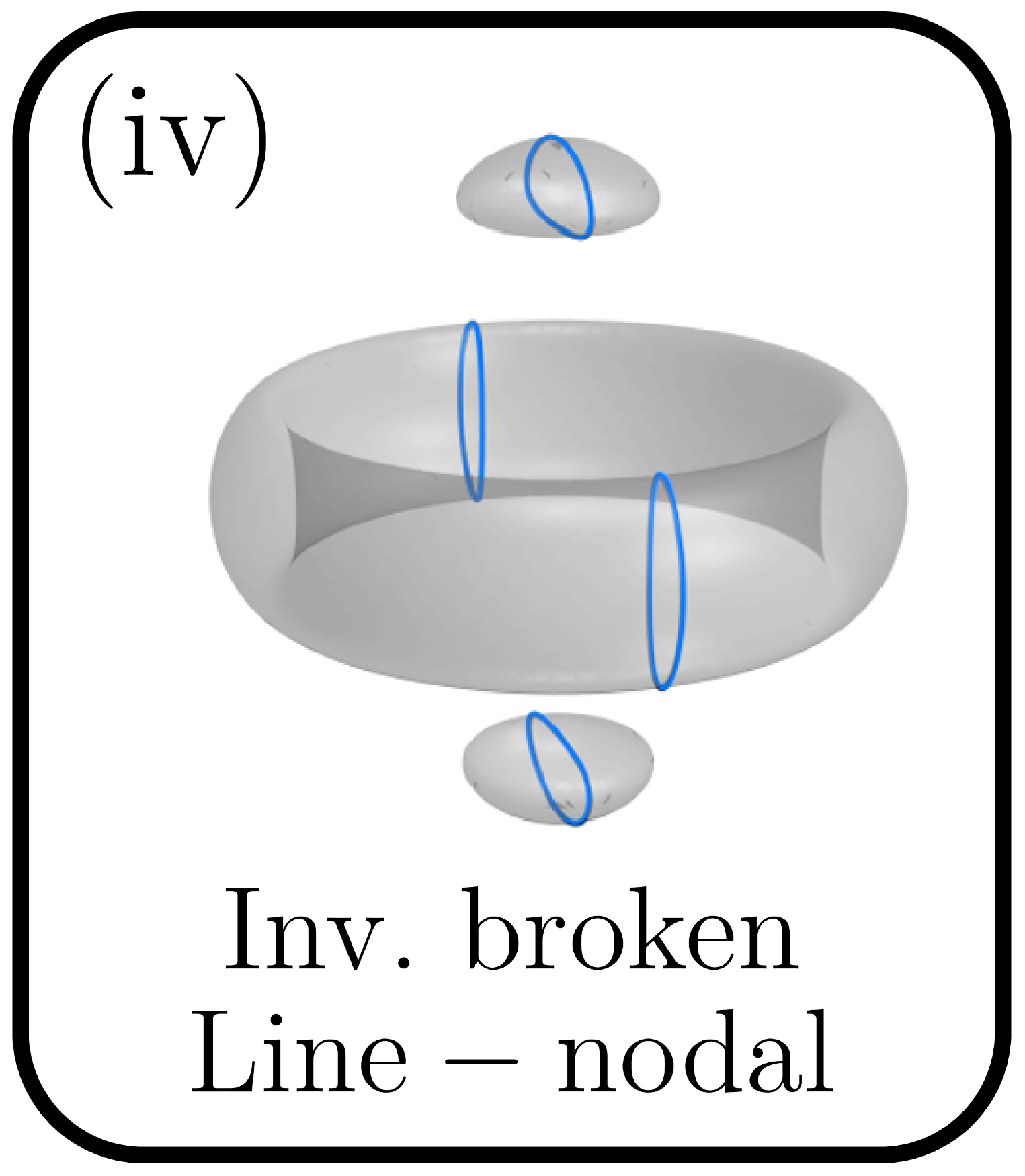} 
\caption{Mean-field phase diagrams at three different temperatures $T_{1} > T_2 > T_3>0$ with two different cases (a,b). 
The corresponding zero-energy manifolds of each phase are illustrated in (c); centrosymmetric BG-FS (i),  a non-centrosymmetric BG-FS (ii,iii), a non-centrosymmetric line-nodal superconductor (iv).   
The phase diagrams are shown for a specific choice of dimensionless parameters, $(g_{x0},T_{1},T_{2},T_{3})/\Lambda=(0.12,10^{-2},10^{-3},10^{-4})$ with a UV-cutoff $\Lambda$.
The ratio between coupling constants, $ g_{xx}/g_{x0} $ and $g_{00}/g_{x0}$ are used for a horizontal and vertical axes, respectively.
For both cases (a,b), centrosymmetric BG-FSs are stable (green, red, blue region) for a weak coupling region and become unstable for a strong coupling limit (white region), where the second-order phase transition boundary is specified by solid lines, Eq.(\ref{se46}). 
As lowering temperatures, the phase space for centrosymmetric BG-FS shrinks and eventually vanishes at zero temperature.
For case 1), a BG-FS survives even after an inversion symmetry breaking (a). 
For case 2), a BG-FS may be gapped and transforms into a line-nodal state after an inversion symmetry breaking, and it dominates over a BG-FS at zero temperature (b).
At dark grey regions $g_{xx}g_{00}<g_{x0}^{2}$, the mean-field free energy is not stable and needs higher-order terms to cure it.  } 
\label{sf3}
\end{center}
\end{figure}

 In Fig. \ref{sf3}, we illustrate schematic phase diagrams at three different temperatures, determined by our phase boundary calculations. Two cases are considered: $\textit{1)}\ \rho_{0}=\hat{k}_{y},\rho_{x}=\hat{k}_{x}$ and $\textit{2})\ \rho_{0}=\rho_{x}=\hat{k}_{x}$. Note that the case $\textit{1)}$ is generic and the case $\textit{2)}$ is fine-tuned. 

\begin{figure}[H]
\begin{center}

\end{center}
\end{figure}

\section{Induced interaction between BG quasiparticles}
In this section, we show how an interaction channel between BG quasiparticles may be related with an interaction channel between electrons. 
Our strategy is to rewrite physical operators of electrons in terms of BG quasi-particles by using symmetries.  

We start with an interaction channel of electrons, 
\begin{eqnarray}
H_{\mathrm{int}}&=& 
 \sum_{\vec{k},\vec{k'}}V_{\mathrm{eff}}(k-k') (\xi_{\vec{k}}^{\dagger}\xi_{\vec{k}'}) (\xi_{-\vec{k}}^{\dagger}\xi_{-\vec{k}'}),
\end{eqnarray}   
with a four-component spinor $\xi_{\vec{k}}^{T}=(f_{\vec{k},\frac{3}{2}},f_{\vec{k},\frac{1}{2}},f_{\vec{k},-\frac{1}{2}},f_{\vec{k},-\frac{3}{2}})$. Note that this type of interaction Hamiltonians include a conventional attractive/repulsive interaction. 
If we assume SO(3) symmetry, then the potential term $V_{\mathrm{eff}}(k-k')$ may be decomposed by spherical harmonics with $(l,m)$,
\begin{eqnarray}
V_{\mathrm{eff}}(k-k') = V_0 + V_1 (\hat{k}\cdot \hat{k}') +\cdots.
\end{eqnarray}

Let us consider the generalized Fierz identity \cite{Lucile2}, 
\begin{eqnarray}
(\xi_{\vec{k}}^{\dagger}\xi_{\vec{k}'})(\xi_{-\vec{k}}^{\dagger}\xi_{-\vec{k}'})=\frac{1}{4}\sum_{S=0}^{3}(\xi_{\vec{k}}^{\dagger}\vec{M}_{S}U_{T}\xi_{-\vec{k}}^{*}) \cdot ( \xi_{-\vec{k}'}^{T}U_{T}^{T}\vec{M}_{S}^{\dagger}\xi_{-\vec{k}'}),
\end{eqnarray}
where $4\times 4$ matrices $U_{T}\equiv i\gamma^{12}$, and $\vec{M}_{S}$ are expressed in Table. \ref{st3}.
We focus on the $l=1$ channel of the potential term, 
which becomes
\begin{eqnarray}
H_{\mathrm{int}}&\rightarrow&\sum_{R'}\frac{V_{1}}{3}\left(\int_{k} \xi_{\vec{k}}^{\dagger}\vec{\delta}_{R'}(\hat{k})U_{T}\xi_{-\vec{k}}^{*} \right)\cdot \left(\int_{k'} \xi_{-\vec{k}'}^{T}U_{T}^{T}\vec{\delta}_{R'}^{\dagger}(\hat{k}')\xi_{\vec{k}'}\right),
\end{eqnarray}
where the $4\times 4$ matrices $\vec{\delta}_{R'}(\hat{k})$ are direct products of $\hat{k}$ and $\vec{M}_{R}(\hat{k})$, and already introduced in Table \ref{st2}. 

Now, one can obtain the interaction term in terms of the four-band and two-band models by projecting the spinors as done in section \ref{s3}, 
\begin{eqnarray}
H_{\mathrm{int}}^{(4)}&=&\sum_{R'}\frac{V_{1}}{3}\left(\int_{k} \psi_{\vec{k}}^{\dagger}\vec{\tilde{\delta}}_{R'}(\hat{k})i\sigma^{y}\psi_{-\vec{k}}^{*} \right)\cdot \left(\int_{k'} \psi_{-\vec{k}'}^{T}(i\sigma^{y})^{T}\vec{\tilde{\delta}}_{R'}^{\dagger}(\hat{k}')\psi_{\vec{k}'}\right),
\end{eqnarray}
\begin{eqnarray}
H_{\mathrm{int}}^{(2)}&=&\sum_{R'}\frac{V_{1}}{3}\left(\int_{k} \Psi_{\vec{k}}^{\dagger}\vec{\rho}_{\mu}^{R'}(\hat{k})\tau^{\mu}\Psi_{-\vec{k}} \right)\cdot \left(\int_{k'} \Psi_{-\vec{k}'}^{\dagger}\vec{\rho}_{\mu}^{R'}(\hat{k}')\tau^{\mu}\Psi_{\vec{k}'}\right).
\end{eqnarray}
where $\vec{\tilde{\delta}}_{R'}(\hat{k}), \vec{\rho}_{\mu}^{R'}(\hat{k})$ are defined in section \ref{s3}. The superscripts, $(4)$ and $(2)$, are to specify the four-band and two-band models. 
Note that the coefficient of the two- and four-models may be also generated by projected other terms. Still, it is possible that the sign of the BG quasiparticles are mainly determined by the bare term $V_1$, and our calculations indicate that the sign of the interaction term between electrons may be inherited to BG quasiparticles. In other words, an attractive interaction between electrons, which may be realized by the conventional electron-phonon mechanism, may give an attractive interaction between BG quasi-particles. 
  \begin{table}[tb]
\centering
\renewcommand{\arraystretch}{1.3}
\begin{tabular}{C{0.06\linewidth}C{0.08\linewidth}C{0.25\linewidth}}\hline \hline

$S$  & $R$ & $\vec{M}_{S}$ (or  $\vec{M}_{R}$ )\\ \hline
$0$  & $A_{1g}$ & $I_{4}$  \\
$1$  & $T_{1g}$ & $\vec{j}$  \\
$2$  & $E_{g}$ & $(\gamma^{4},\gamma^{5})$  \\
	 & $T_{2g}$ & $(\gamma^{1},\gamma^{2},\gamma^{3})$  \\
$3$  & $A_{2g}$ & $\frac{2}{\sqrt{3}}(J_{1}J_{2}J_{3}+J_{3}J_{2}J_{1})$  \\
	 & $T_{1g}$ & $\vec{\mathcal{J}}$  \\
	 & $T_{2g}$ & $\vec{T}$  \\
\hline
\hline
\end{tabular}
\caption{ The spin pairing matrices of $O_{h}$ symmetry. The $4\times 4$ matrices are defined in the caption of Table. \ref{st2}}
\label{st3}
\end{table}

\section{Renormalization Group Analysis}
\begin{figure}[tb]
\begin{center}
\includegraphics[scale=0.8]{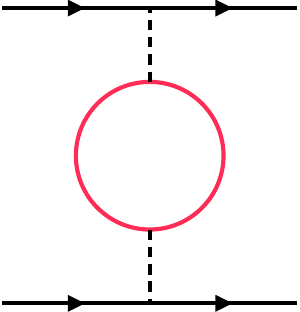}
\caption{A Feynman diagram contributing to the RG flow equations. 
Red line refers to Green's function of a fermion with a fast momentum which is integrated out. We consider a separable four-fermion interaction here, hence there is only one Feynman diagram which contributes to RG equations at one-loop order.  }
\label{sf4}
\end{center}
\end{figure}

We perform the RG analysis on a two-band BdG Hamiltonian,
\begin{eqnarray}
H &= &\sum_{\vec{k}} \Psi^{\dagger}_{\vec{k}}\  E_0(\vec{k}) \tau^z \ \Psi_{\vec{k}}-\frac{1}{2} \sum_{\mu,\nu=0,x} g_{\mu\nu}\sum_{\vec{k},\vec{k}'} (\Psi^{\dagger}_{\vec{k}} \rho_{\mu}(\hat{k})\tau^{\mu} \Psi_{\vec{k}}) ( \Psi^{\dagger}_{\vec{k}'} \rho_{\nu}(\hat{k'})\tau^{\nu} \Psi_{\vec{k}'} ),  \label{se50}\end{eqnarray}
with three coupling constants ($g_{xx},g_{x0},g_{00}$). 
One type of Feynman diagram in Fig. \ref{sf4} contributes to RG equations due to the momentum conservation.

The quantum correction to coupling constants are obtained by integrating out the fast modes, and the coupling constants are modified as $ g_{\mu\nu}\rightarrow  g_{\mu\nu}+\delta g_{\mu\nu}$ with 
\begin{eqnarray}\label{se62}
\delta g_{xx}&=&\ -g_{xx}^{2}\int_{\vec{k},k_n} \mathrm{Tr}\left(G_{f}^{(1)}(\vec{k},k_{n})\tau^{x}G_{f}^{(1)}(\vec{k},k_{n})\tau^{x} \right)\rho_{x}(\hat{k})^2,\nonumber\\
\delta g_{x0}&=&\ -g_{xx}g_{x0}\int_{\vec{k},k_n} \mathrm{Tr}\left(G_{f}^{(1)}(\vec{k},k_{n})\tau^{x}G_{f}^{(1)}(\vec{k},k_{n})\tau^{x} \right)\rho_{x}(\hat{k})^2,\\
\delta g_{00}&=&-g_{x0}^{2}\int_{\vec{k},k_n} \mathrm{Tr}\left(G_{f}^{(1)}(\vec{k},k_{n})\tau^{x}G_{f}^{(1)}(\vec{k},k_{n})\tau^{x} \right)\rho_{x}(\hat{k})^2.\nonumber 
\end{eqnarray}
The energy shell integration, $\int_{\vec{k},k_n} \equiv \int \frac{d^3 k}{(2\pi)^3}\int^{\Lambda}_{-\Lambda}\frac{dk_{n}}{2\pi}$, is used with a high energy cutoff $\Lambda$, which may be an order of a Fermi energy. The bare fermion propagator $G_{f}^{(1)}(\vec{k},k_{n})$ is
\begin{eqnarray}
G_{f}^{(1)}(\vec{k},k_{n})&=&\mathrm{diag}(\frac{1}{-ik_{n}+E_{0}(\vec{k})},\frac{1}{-ik_{n}-E_{0}(\vec{k})}).
\label{se55}
\end{eqnarray}
For example, one of the integrations can be evaluated as 
\begin{eqnarray}
\int_{\vec{k},k_n} \mathrm{Tr}\left(G_{f}^{(1)}(\vec{k},k_{n})\tau^{x}G_{f}^{(1)}(\vec{k},k_{n})\tau^{x} \right)\rho_{x}(\Omega_{\vec{k}})^2&=&
-\int^{\Lambda}_{\Lambda/b} d\epsilon  d\Omega_{\vec{k}} \frac{ N(\epsilon, \Omega_{\vec{k}})\rho_{x}(\Omega_{\vec{k}})^2}{\epsilon},
\end{eqnarray}
with a angle-dependent density of states, $N(\epsilon,\Omega_{\vec{k}})$.

The integration may be further simplified as, 
\begin{eqnarray}
\int^{\Lambda}_{\Lambda/b} d\epsilon  d\Omega_{\vec{k}} \frac{ N(\epsilon, \Omega_{\vec{k}})\rho_{x}(\Omega_{\vec{k}})^2}{\epsilon} &\simeq &\int_{\Lambda/b}^{\Lambda}  \frac{d \epsilon}{\epsilon} \langle N(0,\Omega_{\vec{k}}) \rho_{x}(\Omega_{\vec{k}})^{2} \rangle_{\Omega}=\langle \rho_{x}^2\rangle_{\mathrm{FS}}\log b.
\end{eqnarray}

\subsection{RG equations}
The flow equations of three coupling constants ($b=e^{l}$) are
\begin{eqnarray}
\frac{d \tilde{g}_{xx}}{dl }=\tilde{g}_{xx}^2,\ \ \frac{d \tilde{g}_{x0}}{dl }=\tilde{g}_{x0}\tilde{g}_{xx},\ \ \frac{d \tilde{g}_{00}}{dl }&=& \tilde{g}_{x0}^2,
\end{eqnarray}
with dimensionless coupling constants $\tilde{g}_{\mu\nu}\equiv \langle \rho_{x}^2\rangle_{\mathrm{FS}} \, g_{\mu\nu}$ up to the one-loop calculations. 
We find the analytical solution, 
\begin{eqnarray}
\tilde{g}_{xx}(l)=\frac{\tilde{g}_{xx}(0)}{1-\tilde{g}_{xx}(0)l },\ \ \tilde{g}_{x0}(l)= \frac{\tilde{g}_{x0}(0)}{1-\tilde{g}_{xx}(0)l },\ \ \tilde{g}_{00}(l)&=& \tilde{g}_{00}(0)+\left(\frac{\tilde{g}_{x0}(0)^2 l}{1-\tilde{g}_{xx}(0)l }\right).
\end{eqnarray}
All three coupling constants diverge at the long wavelength limit $l\rightarrow l_{c}\equiv \tilde{g}_{xx}(0)^{-1}$, which indicates instability of a BG-FS. 
 
 Our results may be generalized by including additional coupling constants. We find the flow equations with  six coupling constants, 
\begin{eqnarray}
\frac{dg_{xx}}{dl }=g_{xx}^2\langle \rho_{x}^{2}\rangle_{\mathrm{FS}}+g_{xy}^2\langle \rho_{y}^2\rangle_{\mathrm{FS}},\ \ \ \ \ &\ \ \ &\frac{d g_{yy}}{dl }=g_{xy}^2\langle \rho_{x}^2\rangle_{\mathrm{FS}}+g_{yy}^2\langle \rho_{y}^2\rangle_{\mathrm{FS}},\nonumber\\
\frac{d g_{xy}}{dl }= g_{xx}g_{xy}\langle \rho_{x}^2\rangle_{\mathrm{FS}}+g_{yy}g_{xy}\langle \rho_{y}^2\rangle_{\mathrm{FS}},&\ \ \ &
\frac{d g_{x0}}{dl }=g_{x0}g_{xx}\langle \rho_{x}^2\rangle_{\mathrm{FS}}+g_{y0}g_{xy}\langle \rho_{y}^2\rangle_{\mathrm{FS}}, \\
\frac{d g_{y0}}{dl }=g_{x0}g_{xy}\langle \rho_{x}^2\rangle_{\mathrm{FS}}+g_{y0}g_{yy}\langle \rho_{y}^2\rangle_{\mathrm{FS}}&\ ,\ &\frac{d g_{00}}{dl }= g_{x0}^2\langle \rho_{x}\rangle_{\mathrm{FS}}^2+ g_{y0}^2\langle \rho_{y}\rangle_{\mathrm{FS}}^2.\nonumber
\end{eqnarray}
We check that the RG flows are away from the non-interacting fixed point in the long wavelength limit $l\rightarrow \infty$, and hence an instability of inversion symmetry exists. 
 
Furthermore, we may consider repulsive interactions by considering negative values of $(g_{xx},g_{x0},g_{00})$.
We find that the flow equation goes into another non-trivial fixed point $(\tilde{g}_{xx}^{*},\tilde{g}_{x0}^{*},\tilde{g}_{00}^{*})=(0,0,\tilde{g}_{00}(0) - \frac{\tilde{g}_{x0}(0)^2 }{\tilde{g}_{xx}(0)})$. 
The fixed point is reliable when the coupling constants at the fixed point are small, so we consider the region where $ \tilde{g}_{x0}(0)^2   \ll \tilde{g}_{00}(0)\tilde{g}_{xx}(0)$.
Then, one can conclude that a BG-FS is stable with marginal interaction $\tilde{g}_{00}^{*}$ upto the one-loop calculation. We note that the $\tilde{g}_{00}(0)$ channel is a type of density-density interaction in terms of $\Psi_k$, so we conclude it does not induce an instability of a BG-FS even in higher orders if $\tilde{g}_{00}(0) \ll 1$.

%

  \end{document}